\documentclass[pdflatex,sn-mathphys-num]{sn-jnl}

\usepackage{graphicx}%
\usepackage{multirow}%
\usepackage{amsmath,amssymb,amsfonts}%
\usepackage{amsthm}%
\usepackage{mathrsfs}%
\usepackage[title]{appendix}%
\usepackage{xcolor}%
\usepackage{textcomp}%
\usepackage{manyfoot}%
\usepackage{booktabs}%
\usepackage{algorithm}%
\usepackage{algorithmicx}%
\usepackage{algpseudocode}%
\usepackage{listings}%
\usepackage{physics}

\newcommand{\mb}[1]{\mathbf{#1}}

\newcommand{\what}[1]{\widehat{#1}}
\newcommand{\oline}[1]{\overline{#1}}
\renewcommand{\ip}[2]{\left\langle#1,#2 \right\rangle}
\newcommand{\mc}[1]{\mathcal{#1}}
\newcommand{\bsb}[1]{\boldsymbol{#1}}

\newcommand{\vbar}[1]{\overline{\mathbf{#1}}}
\newcommand{\vhat}[1]{\widehat{\mathbf{#1}}}

\renewcommand{\vb}[1]{\boldsymbol{#1}}

\renewcommand{\vbar}[1]{\overline{\boldsymbol{#1}}}
\renewcommand{\vhat}[1]{\widehat{\boldsymbol{#1}}}

\definecolor{PrlBlue}{HTML}{2e3092}
\definecolor{CTI_ORANGE_SRG}{rgb}{0.996, 0.421, 0.046}
\definecolor{RED}{rgb}{0.546875,    0.08203125,    0.08203125}
\definecolor{BLUE}{rgb}{0.0,    0.328125,    0.55859375}
\definecolor{GREEN}{rgb}{0.3020,    0.6863,    0.2902}
\definecolor{PURPLE}{rgb}{0.5961,    0.3059,    0.6392}
\definecolor{ORANGE}{rgb}{1.0    0.4980    0.0}

\definecolor{YELLOW}{rgb}{1.0000,    1.0000,    0.2000}
\definecolor{BROWN}{rgb}{0.6510,    0.3373,    0.1569}
\definecolor{PINK}{rgb}{0.9686    0.5059    0.7490}
\definecolor{CYAN}{rgb}{0,    0.9961,    0.9961}
\definecolor{NEON}{rgb}{0.2227,    0.9961,    0.0781}
\definecolor{GREY}{rgb}{0.4500,    0.4500,    0.3500}

\newcommand\Real{\mbox{Re}}          

\theoremstyle{thmstyleone}%
\theoremstyle{thmstyletwo}%

\theoremstyle{thmstylethree}%

\usepackage{listings}
\lstdefinestyle{MatlabStyle}{
  language=Matlab,
  basicstyle=\ttfamily\small,
  keywordstyle=\color{blue},
  stringstyle=\color{orange},
  commentstyle=\color{green!50!black},
  numbers=left,
  numberstyle=\tiny\color{gray},
  stepnumber=1,
  numbersep=5pt,
  showstringspaces=false,
  breaklines=true,
  frame=single,
  rulecolor=\color{black},
  tabsize=2,
  xleftmargin=0pt,
  framexleftmargin=0pt,
  framextopmargin=5pt,
  framexbottommargin=5pt,
  captionpos=b
}

\begin{document}

\title[Article Title]{Uncertainty Quantification in Resolvent Analysis of Experimental Wall-Bounded Turbulent Flows}


\author*[1]{\fnm{Salvador R.} \sur{Gomez}}\email{gomezsr@stanford.edu}

\author[1]{\fnm{Tomek M.} \sur{Jaroslawski}}\email{tomek@stanford.edu}
\equalcont{These authors contributed equally to this work.}

\affil*[1]{\orgdiv{Center for Turbulence Research}, \orgname{Stanford University}, \orgaddress{\street{Building 500 Escondido Mall}, \city{Stanford}, \postcode{94305}, \state{California}, \country{United States of America}}}


\abstract{
Experimental mean flows are commonly used to study wall-bounded turbulence. However, these measurements are often unable to resolve the near-wall region and thus introduce ambiguity in the velocity closest to the wall. This poses a source of uncertainty in equation-based approaches that rely on these mean flow measurements such as resolvent analysis. Resolvent analysis provides a scale-dependent decomposition of the linearized Navier Stokes equations that identifies optimal gains, response modes and forcing modes that has been used to great effect in turbulent wall-bounded flows. Its potential in the development of predictive tools for a variety of wall-bounded flows is high but the limitations of the input data must be addressed. Here, we quantify the sensitivity of resolvent analysis to common sources of experimental uncertainty and show that this sensitivity can be quantified with minimal additional computational cost. This approach is applied to both local and biglobal resolvent analysis by using artificial disturbances to mean profiles in the former and particle image velocimetry measurements with differing near-wall fits in the latter. We also highlight an example where poor near-wall resolution can lead to erroneous conclusions compared to the full-resolution data in an adverse pressure gradient turbulent boundary layer.}




\maketitle

\section{Introduction}\label{Sec_Intro}

    Wall-bounded turbulent flows exhibit strong inhomogeneity and a wide range of interacting scales, making them particularly difficult to measure and characterize. Resolvent analysis provides a linear framework to study the amplification of disturbances by the mean flow, but its accuracy depends critically on the fidelity of that mean profile. In practice, experimental measurements often lack sufficient resolution close to the wall. Similarly, wall-modeled large-eddy simulations (WMLES) rely on models to approximate near-wall dynamics, introducing significant uncertainty in that region. As a result, mean flow profiles may contain inaccuracies—especially near the wall—raising an important question: how sensitive is resolvent analysis to such errors? This study addresses that question through a systematic sensitivity analysis.

    The mean flow field of a zero-pressure-gradient (ZPG) turbulent boundary layer (TBL) on a hydraulically smooth surface exhibits self-similar behavior across different wall-normal regions, each with its own characteristic length and velocity scales. These scales give rise to the classical law of the wall and the law of the wake, which are valid in the near-wall and outer regions of the flow, respectively, with the log layer classically characterized as an overlap layer where both scaling laws apply~\citep{coles1956law}. These scaling laws can be used to model the near-wall flow field from outer region measurements~\citep{van1956turbulent}. The near-wall scales are determined by viscosity and wall shear via the friction velocity, $u_{\tau} = \sqrt{\tau_{w}/\rho}$, where $\tau_{w}$ is the wall shear stress and $\rho$ the density, and the viscous length scale, $\ell_{\nu} = \nu/u_{\tau}$, where $\nu$ is the kinematic viscosity. Scales normalized with $u_{\tau}$ and $\ell_{\nu}$ are here denoted with a $+$ superscript. The outer region can be described by the outer length-scale, $\delta$. A measure of scale separation in a ZPG TBL is the friction Reynolds number, $Re_{\tau} = \delta/\ell_{\nu}$.

    The importance of the mean flow field is that it encodes valuable information about the dynamics of the turbulent fluctuations through turbulent production and linear amplification. The linearized Navier Stokes equations (LNSE) create a linear operator, $\mb{L}$, that can provide many useful predictions for turbulent fluctuations. Linear stability analysis is used to determine the growth of initial disturbances~\citep{schmid2002stability}, though these results are sensitive to perturbations in $\mb{L}$ when $\mb{L}$ is non-normal~\citep{trefethen1993hydrodynamic,reddy1993pseudospectra}, \textit{i.e.} $\mb{L}\mb{L}^{\dagger}\ne\mb{L}^{\dagger}\mb{L}$. Transient growth analysis relies on the non-normality of $\mb{L}$~\citep{trefethen1993hydrodynamic} and has identified length-scales and flow structures similar to those observed in turbulent flows from the optimal linear amplification~\citep{del2006linear,cossu2009optimal,hwang2010linear}. While these approaches can augment $\mb{L}$ with an eddy viscosity to model the nonlinearities~\citep{reynolds1972mechanics}, they ultimately ignore the presence of the nonlinear terms in the fluctuating equations. Resolvent analysis remedies this by modeling the nonlinearities as a forcing input and creating the resolvent operator as a linear transfer function between the fluctuations and said forcing~\citep{mckeonsharma2010}. The singular value decomposition (SVD) of the resolvent operator identifies an orthonormal forcing basis and an orthonormal response basis each ranked by their singular values, the linear gain of the transfer function. With deep ties to the psuedospectrum of $\mb{L}$, resolvent analysis takes advantage of non-normal mechanisms present to find the most amplified motions~\citep{symon2018non}. Resolvent analysis has identified  similarities in the response modes to those observed in the near-wall turbulent flows~\citep{moarref2013model,abreu2020spectral}, increased amplification of the large-scales similar to the observed increased energization in APG TBLs~\citep{gomez2025linearAPGTBL}, provides a useful basis for flow reconstruction~\citep{moarref2014low_order_represenanation}, and can agree with data-driven modal decompositions~\citep{towne2018spectral}. 
    
    Extensions to high Reynolds number wall-bounded flows, especially those with non-canonical effects like pressure gradients or roughness, are limited by the availability of well-resolved data and often rely on experimental measurements. Experimental mean flow fields have been used in resolvent analysis to great success~\citep{beneddine2017unsteady,he2019data,lesshafft2019resolvent,symon2019tale,preskett2024CTRSP}, and we highlight the observations of \citet{chavarin2020resolvent} that their resolvent gains were sensitive to the mean flow estimation. However, experimental measurements in high Reynolds number flows often have a great deal of uncertainty near the wall where the law-of-the-wall is often used to supplement the lack of measurements in this region. The sensitivity of linear stability analysis to experimental measurements and fitting of the mean profile has been studied by \citet{boutilier2013sensitivity} and can be rigorously quantified with psuedospectral methods~\citep{reddy1993pseudospectra,trefethen1999spectra}. The sensitivity of resolvent gains to parameters has been used to great effect to study parameterization of the mean flow and the LNSE~\citep{de2014parametric,fosas2017optimal,skene2019adjoint} and to search large amplification regions in wavenumber space~\citep{gomezCTRSP2022}. Here, we use this approach to quantify the uncertainty in the resolvent analysis gains to experimental mean flow error and mean flow reconstruction. Furthermore, we extend the approaches in \citet{de2014parametric,fosas2017optimal,skene2019adjoint,gomezCTRSP2022} to introduce a formulation that predicts the sensitivity in the resolvent modes as well. In particular, we focus on uncertainty in the near-wall measurements since the near-wall mean shear is a source of amplification responsible for the low-rank behavior in the resolvent operator that leads to agreement with observed turbulent motions~\citep{moarref2013model,symon2018non,abreu2020spectral}.

    Near-wall velocity measurement remains the dominant experimental uncertainty in turbulent boundary layers. Standard techniques like particle image velocimetry (PIV), laser Doppler anemometry (LDA), and hot-wire anemometry (HWA) are limited by spatial resolution, optical access, wall interference, and steep gradients; these limitations intensify at high Reynolds number and in thin boundary layers. Advanced methods exist, but their complexity and limited availability restrict routine use, so near-wall velocity is often poorly constrained in typical laboratory setups.
    
    Hot-wire anemometry near the wall is affected by finite probe size, sensor-induced flow disturbance, and heat conduction to the wall \citep{hutchins2009hot,durst2001situ,orlu2010near}. Accurate wall-location determination is critical yet difficult: reported accuracies range from about 25–200 µm depending on probe size and plate material \citep{bruun1996hot,ligrani1987spatial}, with improvements via electrical circuits \citep{bhatia1982corrections}, microscopy \citep{hutchins2009hot}, and reflected-image techniques \citep{orlu2010near}; best cases report uncertainty as low as ±5 µm. Consequently, mean velocities within the viscous sublayer are generally unreliable unless stringent procedures are followed.
    
    The friction velocity, $u_{\tau}$, is commonly inferred from log-layer fits (e.g., Clauser plots), which are not direct $\tau_{w}$ measurements. Typical fitting errors are about 5\% \citep{clauser1956turbulent}; multi-parameter fits (Spalding/Musker-type) yield similar magnitudes \citep{musker1979explicit}. In non-equilibrium conditions, such as adverse pressure gradients, errors can rise to ~10\% \citep{dixit2009determination}. Direct $\tau_{w}$ measurements (oil-film interferometry, MEMS shear sensors, Preston tubes) each carry method-specific uncertainties and reduced reliability in three-dimensional or unsteady flows. Reported uncertainties for MEMS fences, Preston tubes, wall-mounted hot wires, and pulsed-wire methods are typically on the order of 4\%, while oil-film interferometry can achieve $<4\%$ with stringent control of temperature, viscosity calibration, and imaging \citep{nagib2007approach,fernholz1996new,patel1965calibration}. 
    
    PIV provides full-field velocity but remains challenged near the wall by reflections, low seeding, limited resolution, particle-image bias, and window-correlation bias; near-wall data are often omitted or unreliable, and the viscous sublayer is typically not resolved \citep{kahler2012uncertainty}. Consequently, velocity-profile fitting or extrapolation is required, and associated errors must be propagated in interpretation. While advanced near-wall PIV/HWA methods improve performance \citep{fuchs2023wall,huck2025near}, complex configurations (e.g., rough-wall flows) still require some form of fitting. Similarly, RANS and WMLES exhibit significant near-wall modeling uncertainty, affecting velocity gradients, $\tau_{w}$, and heat transfer predictions. State of the art WMLES typically report uncertainties between $10-14\%$ in the skin friction~\citep{piomelli2002wall,kawai2012wall}, translating to $5-7\%$ errors in $u_{\tau}$. The application of these simulations to non-equilibrium conditions can lead to additional error due the use of wall-models based on classical scaling arguments applicable to the ZPG TBL, though extensions to handle non-equilibrium conditions have been developed~\citep{park2014improved}. All these measurement errors contribute to unreliable mean flow fields that then affect the LNSE.

    In this paper, we first derive the resolvent analysis sensitivity for a general flow, describe the artificial near-wall mean flow errors, and describe the PIV set-up in section \ref{Sec_Meth}. The sensitivity analysis is shown to predict the error in local resolvent analysis when errors in the mean up to $45\%$ are artificially introduced in section \ref{Sec_Res1D}. Section \ref{Sec_Res2D} then quantifies the sensitivity of biglobal resolvent analysis using a PIV mean flow field where different near-wall fits are used. In Section \ref{Sec_Implications} we illustrate that a lack of near-wall resolution in APG data leads to erroneous conclusions compared to the fully resolved data and provide suggestions for those using experimental data for resolvent analysis. Finally, conclusions are provided in section \ref{Sec_Conc}.  

\section{Methodology}\label{Sec_Meth}
    Here, we first begin by discussing the resolvent analysis framework and numerical scheme before introducing the sensitivity analysis. We then describe the experimental mean flow field and the artificially introduced near-wall errors along with the near-wall fits to the PIV data. For brevity, we will focus the resolvent sensitivity approach to incompressible flows with mean flow uncertainty within the text. The general sensitivity analysis that is applicable to compressible flows, including uncertainty in the terms of the inner-product weights is presented in Appendix \ref{App}. 

\subsection{Resolvent analysis}\label{Sec_MethRA}
    We first describe the resolvent analysis framework for an incompressible flow. Here, the velocity is $\vb{U} = U\vb{e}_{x} + V \vb{e}_{y} + W \vb{e}_{z}$ with $U$, $V$, and $W$ denoting streamwise, wall-normal, and spanwise components, the spatial coordinates are $\vb{x} = x\vb{e}_{x} + y \vb{e}_{y} + z \vb{e}_{z}$ with $x$, $y$, and $z$ denoting streamwise, wall-normal, and spanwise coordinates, $P$ is the pressure, $\nu$ is the kinematic viscosity, and $t$ is time. The coordinates are separated into homogenous coordinates, $\vb{x}_{h}$, and inhomogenous coordinates, $\vb{x}_{n}$. In this study, we consider the parallel flow assumption where $\vb{x}_{h} = x\vb{e}_{x} + z\vb{e}_{z}$ and $\vb{x}_{n} = y\vb{e}_{y}$ (local analysis) and nonparallel flows where $\vb{x}_{h} = z\vb{e}_{z}$ and $\vb{x}_{n} = x\vb{e}_{x} + y\vb{e}_{y}$ (biglobal analysis). The flow state, $\vb{Q}(\vb{x},t) = \qty[\vb{U},P]^{T}$, is statistically stationary with the observed mean state, $\vbar{Q}(\vb{x}_{n};\vb{a}) = \qty[\vbar{U}(\vb{x}_{n};\vb{a}),\oline{P}(\vb{x}_{n};\vb{a})]^{T}$, and fluctuation $\vb{q}(\vb{x},t) = \qty[\vb{u},p]^{T}$ where $\vbar{Q}$ can depend on parameters $\vb{a}$. While $\vb{a}$ usually includes flow conditions like nondimensional numbers, we use $\vb{a}$ to quantify sources of measurement uncertainty like near-wall fitting coefficients or the absolute position of the wall-normal measurement locations. 

    The evolution of $\vb{q}$ is described by the Navier Stokes equations (NSE), here written as 
    	\begin{equation}
    		\qty(\mb{B}\partial_{t} + \mb{L}(\vbar{Q};\vb{a}))\vb{q} = \mb{B}\vb{n} \label{Meth_GovEqNSE}
    	\end{equation}
    where in an incompressible flow, $\mb{B}$ extracts the velocity as $\mb{B}\vb{q} = \qty[\vb{u},0]^{T}$, $\mb{L}$ is the Jacobian of the NSE evaluated at $\vbar{Q}$ which can also depend on parameters $\vb{a}$, and $\vb{n}$ are the nonlinear terms. Equation \ref{Meth_GovEqNSE} can be Fourier transformed in $\vb{x}_{h}$ and $t$ with Fourier modes denoted with $\what{\cdot}$ such that
    	\begin{equation}
    		\vhat{q}(\vb{x}_{n};\vb{k},\omega) = \int\int\vb{q}(\vb{x},t)e^{i\qty(\vb{k}\cdot\vb{x}_{h}-\omega t)}dt d\vb{x}_{h},
    	\end{equation}  
    where $\omega$ and $\vb{k}$ are the temporal frequency and wavenumber vector. Following \citet{mckeonsharma2010}, $\vb{n}$ will now be treated as a forcing $\vb{f}$ that is uncorrelated from $\vb{q}$ such that the Fourier transformed equations are now
    	\begin{equation}
    		\vhat{q} = \qty(-i\mb{B}\omega + \vhat{L}(\vbar{Q};\vb{a},\vb{k}))^{-1}\vb{B}\vhat{f} =\mb{A}(\vbar{Q};\vb{a},\vb{k},\omega)^{-1}\mb{B}\vhat{f} =  \mb{H}(\vbar{Q};\vb{a},\vb{k},\omega)\vhat{f} \label{Meth_eq_FTNSE}
    	\end{equation}
    where $\mb{A}$ is the LNSE, $\mb{H}$ the resolvent operator, and for incompressible flow, 
    	\begin{equation}
    		\vhat{L} = \mqty[\vbar{U}\cdot\what{\nabla} + \nabla\vbar{U} - \nu \what{\nabla}^{2}&& \what{\nabla} \\ \what{\nabla} \cdot && 0].
    	\end{equation}
    For parallel wall-bounded flows $\vbar{U}(y;\vb{a}) = \oline{U}(y;\vb{a})\vb{e}_{x}$ and $\what{\nabla} = ik_{x}\vb{e}_{x} + \partial_{y}\vb{e}_{y} + ik_{z}\vb{e}_{z}$ while for streamwise developing wall-bounded flows, $\vbar{U}(x,y;\vb{a}) = \oline{U}(x,y;\vb{a})\vb{e}_{x} + \oline{V}(x,y;\vb{a})\vb{e}_{y}$ and $\what{\nabla} = \partial_{x}\vb{e}_{x} + \partial_{y}\vb{e}_{y} + ik_{z}\vb{e}_{z}$.

    We now introduce the inner product, 
    	\begin{equation}
    		\ip{\vb{q}_{1}}{\vb{q}_{2}} = \int \vb{q}_{1}^{*}\mb{B}\vb{q}_{2}d\vb{x}_{n}, \label{eq_inner_def}
    	\end{equation}
    where $*$ denotes a complex conjugate and the norm $\norm{\vb{q}_{1}}^{2} = \ip{\vb{q}_{1}}{\vb{q}_{1}}$ is the kinetic energy norm. We also define the response and forcing inner products as $\ip{\vb{q}_{1}}{\vb{q}_{2}}_{r} = \ip{\vb{q}_{1}}{\mb{W}_{r}(\vb{x}_{n};\vb{a})\vb{q}_{2}}$ and $\ip{\vb{q}_{1}}{\vb{q}_{2}}_{f} = \ip{\vb{q}_{1}}{\mb{W}_{f}(\vb{x}_{n};\vb{a})\vb{q}_{2}}$, respectively, where both $\mb{W}_{r}$ and $\mb{W}_{f}$ are positive semi-definite operators which can have spatial dependence. We will consider the $\vb{a}$ dependence on $\mb{W}_{r}$ and $\mb{W}_{f}$ in Appendix \ref{App}. The adjoint operators with respect to the inner product in equation \ref{eq_inner_def} are denoted with a $\dagger$.

    In resolvent analysis, one seeks the $\bsb{\phi}$ that produces the largest $\ip{\mb{H}\bsb{\phi}}{\mb{H}\bsb{\phi}}_{r}$ such that $\ip{\bsb{\phi}}{\bsb{\phi}}_{f} = 1$. It can be shown that this optimization can be solved through the eigenvalue problem, 
        \begin{equation}
            \mb{H}^{\dagger}\mb{W}_{r}\mb{H}\bsb{\phi}_{i} = \sigma_{i}^{2}\mb{W}_{f}\bsb{\phi}_{i}. \label{Eq_EVP_SVD}
        \end{equation}
    This leads to a decomposition of $\mb{H}$ as $\mb{H} = \sum_{i}\sigma_{i}\bsb{\psi}_{i}\bsb{\phi}_{i}^{*}$ where $\bsb{\psi}_{i}$ are the orthonormal response modes, $\bsb{\phi}_{i}$ the orthonormal forcing modes, and $\sigma_{i}$ the linear gains where the indices $i$ are ordered such that $\sigma_{i} \ge \sigma_{i+1}$. The response modes and forcing modes are related by
        \begin{equation}
            \mb{H}\bsb{\phi}_{i} = \sigma_{i}\bsb{\psi}_{i} \label{Eq_MainRAEquation}
        \end{equation}
    and
        \begin{equation}
            \mb{W}_{f}^{-1}\mb{H}^{\dagger}\mb{W}_{r}\bsb{\psi}_{i} = \sigma_{i}\bsb{\phi}_{i} \label{Eq_MainRAEquationAdj}
        \end{equation}
    as properties of the SVD and the choice of inner products. 

    The numerical schemes follow \citet{gomez2024linear} and \citet{gomez2025linearAPGTBL}. The equations are discretized using a fourth-order summation by parts scheme~\citep{mattsson2004SBP} where the streamwise direction is equispaced and grid stretching is employed in the wall-normal direction so that half the points lie in $[0,y_{min})$ and the other half lie in $(y_{min},y_{max}]$ as in \citet{malik1990numerical}. For the local analysis, we use $N_{y}=251$ points and set $y_{min}^{+} = 100$ and $y_{max}/\delta = 3$. The biglobal analysis uses $N_{x} = 250$ points in the streamwise direction and the same wall-normal discretization as the local analysis. See \citet{gomez2024linear} for the algorithm used to solve the eigenvalue problem in equation \ref{Eq_EVP_SVD}.

    \subsection{Sensitivity analysis}\label{Sec_MethSA}
        
        Here, we will present the sensitivity analysis of $\sigma_{i}$ similar to the one presented in \citet{de2014parametric}, \citet{fosas2017optimal}, and \citet{skene2019adjoint}. In addition, we will quantify the perturbations of $\bsb{\psi}_{i}$ and $\bsb{\phi}_{i}$. First, differential forms of Equations \ref{Eq_MainRAEquation}, \ref{Eq_MainRAEquationAdj}, and the orthonormality constraints of $\bsb{\psi}_{i}$ and $\bsb{\phi}_{i}$ are introduced as 
            \begin{align}
                    \partial\mb{H}\bsb{\phi}_{i} + \mb{H}\partial\bsb{\phi}_{i} &= \partial \sigma_{i} \bsb{\psi}_{i} + \sigma_{i}\partial\bsb{\psi}_{i},\label{Eq_Diff_MainRAEq}\\
                    \partial\mb{H}^{\dagger}\mb{W}_{r}\bsb{\psi}_{i} + \mb{H}^{\dagger}\partial\mb{W}_{r}\bsb{\psi}_{i} + \mb{H}^{\dagger}\mb{W}_{r}\partial\bsb{\psi}_{i} &= \partial\sigma_{i}\mb{W}_{f}\bsb{\phi}_{i} + \sigma_{i}\partial\mb{W}_{f}\bsb{\phi}_{i} + \sigma_{i}\mb{W}_{f}\partial\bsb{\phi}_{i},\label{Eq_Diff_MainRAEqAdj}\\
                    \ip{\partial\bsb{\psi}_{j}}{\mb{W}_{r}\bsb{\psi}_{i}} + \ip{\bsb{\psi}_{j}}{\mb{W}_{r}\partial\bsb{\psi}_{i}} &= -\ip{\bsb{\psi}_{j}}{\partial\mb{W}_{r}\bsb{\psi}_{i}}, \label{Eq_Diff_PsiResp}\\
                    \ip{\partial\bsb{\phi}_{j}}{\mb{W}_{f}\bsb{\phi}_{i}} + \ip{\bsb{\phi}_{j}}{\mb{W}_{f}\partial\bsb{\phi}_{i}} &= -\ip{\bsb{\phi}_{j}}{\partial\mb{W}_{f}\bsb{\phi}_{i}}.\label{Eq_Diff_PhiResp}
                \end{align}
        respectively. The differential forms of $\mb{H}$ and $\mb{H}^{\dagger}$ are $\partial \mb{H} = -\mb{H}\partial\mb{A}\mb{H}$ and $\mb{H}^{\dagger} = -\mb{H}^{\dagger}\partial\mb{A}^{\dagger}\mb{H}^{\dagger}$. Hereafter, we will focus only on mean flow sensitivity which only affects the LNSE and will thus set $\partial\mb{W}_{r} = \partial\mb{W}_{f} = \mb{0}$. See Appendix \ref{App} for a complete description that considers the sensitivity to the parameterization of the LNSE, $\mb{W}_{r}$, and $\mb{W}_{f}$.

        In this analysis, the perturbations stem from near-wall uncertainty affecting only $\vbar{U}$. Thus $\partial\mb{A} = \partial \mb{L}$ where
            \begin{equation}
                \partial\mb{L} = \mqty[ \partial \vbar{U} \cdot \vhat{\nabla} + \grad 
                \partial \vbar{U} & 0 \\  0 & 0].
            \end{equation}
        The perturbation in the LNSE from a change in the mean flow field thus includes a perturbation in the mean convection and the mean shear which create perturbations in the common routes of amplification in resolvent analysis, i.e. the critical layer amplification~\citep{mckeonsharma2010} and the non-normal component-wise amplification~\citep{symon2018non}.

        By taking the $\mb{W}_{r}$-weighted inner product of Equation \ref{Eq_Diff_MainRAEq} with $\bsb{\psi_{i}}$ and making use of equations \ref{Eq_Diff_PsiResp} and \ref{Eq_Diff_PhiResp}, it can be shown that the perturbation of $\sigma_{i}$ is
            \begin{equation}
                \partial\sigma_{i} = -\sigma_{i}^{2}\Real(\ip{\bsb{\phi}_{i}}{\mb{W}_{f}\partial \mb{L} \bsb{\psi}_{i}}), \label{Eq_SimpSensitivitySigma}
            \end{equation}
        where $\Real(\cdot)$ is the real part of a complex number. The perturbation in $\bsb{\psi}_{i}$ can be expressed as \begin{equation}
                \partial\bsb{\psi}_{i} = \sum_{j=1}a_{j,i}\bsb{\psi}_{j}, \label{Eq_Sum_Delta}
            \end{equation}
        where $a_{j,i} = \ip{\bsb{\psi}_{j}}{\mb{W}_{r}\partial\bsb{\psi}_{i}}$ for $j \ne i$. This restriction is placed because in the absence of perturbations to $\mb{W}_{r}$, $\ip{\bsb{\psi}_{i}}{\mb{W}_{r}\partial\bsb{\psi}_{i}} = 0$ from Equation \ref{Eq_Diff_PsiResp}. Instead, $a_{i,i}$ should be determined such that $\bsb{\psi}_{i} + \partial\bsb{\psi}_{i}$ has unit-norm which depends on all of the higher order modes. Here, the perturbation of $\bsb{\psi}_{i}$ will be measured by $\ip{\bsb{\psi}_{j}}{\mb{W}_{r}\partial\bsb{\psi}_{i}}$. By taking the $\mb{W}_{r}$-weighted inner product of Equation \ref{Eq_Diff_MainRAEq} with $\bsb{\psi_{j}}$ and using equations \ref{Eq_Diff_MainRAEqAdj}--\ref{Eq_Diff_PhiResp}, 
            \begin{equation}
                \begin{split}
                \ip{\bsb{\psi}_{j}}{\mb{W}_{r}\partial\bsb{\psi}_{i}} = \frac{1}{\sigma_{i}^{2}-\sigma_{j}^{2}}\left[-\sigma_{i}^{2}\sigma_{j}\ip{\bsb{\phi}_{j}}{\mb{W}_{f}\partial\mb{L}\bsb{\psi}_{i}}-\sigma_{j}^{2}\sigma_{i}\ip{\partial\mb{L}\bsb{\psi}_{j}}{\mb{W}_{f}\bsb{\phi}_{i}}\right]. \label{Eq_SimpSensitivityPsi}
                \end{split}
            \end{equation}
        Similarly, an expression for $\ip{\bsb{\psi}_{j}}{\mb{W}_{r}\partial\bsb{\psi}_{i}}$ can be found by starting with the $\mb{W}_{f}$-weighted inner product of Equation \ref{Eq_Diff_MainRAEqAdj} with $\bsb{\phi_{j}}$ as
            \begin{equation}
                \begin{split}
                \ip{\bsb{\phi}_{j}}{\mb{W}_{f}\partial\bsb{\phi}_{i}} = \frac{1}{\sigma_{i}^{2}-\sigma_{j}^{2}}\left[-\sigma_{i}^{2}\sigma_{j}\ip{\partial\mb{L}\bsb{\psi}_{j}}{\mb{W}_{f}\bsb{\phi}_{i}}-\sigma_{j}^{2}\sigma_{i}\ip{\bsb{\phi}_{j}}{\mb{W}_{f}\partial\mb{L}\bsb{\psi}_{i}}\right]. \label{Eq_SimpSensitivityPhi}
                \end{split}
            \end{equation}
        These expressions introduce the symmetries, $\ip{\bsb{\psi}_{j}}{\mb{W}_{r}\partial\bsb{\psi}_{i}} = -\ip{\bsb{\psi}_{i}}{\mb{W}_{r}\partial\bsb{\psi}_{j}}^{*}$, $\ip{\bsb{\phi}_{j}}{\mb{W}_{f}\partial\bsb{\phi}_{i}} = -\ip{\bsb{\phi}_{i}}{\mb{W}_{f}\partial\bsb{\phi}_{j}}^{*}$, and when $\mb{W}_{r} = \mb{W}_{f}$, $\ip{\bsb{\psi}_{j}}{\mb{W}_{r}\partial\bsb{\psi}_{i}} = \ip{\bsb{\phi}_{j}}{\mb{W}_{f}\partial\bsb{\phi}_{i}}$.

        Together, equations \ref{Eq_SimpSensitivitySigma}--\ref{Eq_SimpSensitivityPhi} predict the perturbations to $\sigma_{i}$, $\bsb{\psi}_{i}$, and $\bsb{\phi}_{i}$ as a function of only $\partial\mb{L}$ and the computed resolvent modes and gains. Notably, these predictions avoid the use $\mb{H}$ and $\mb{H}^{\dagger}$, relying only on matrix multiplications and inner products of already computed quantities. Thus, the sensitivity of the resolvent analysis to known perturbations can be computed with a negligible computational cost involving only matrix multiplications. Furthermore, $\partial\mb{L}$ acts locally on $\bsb{\psi}_{i}$ meaning that modes are sensitive to perturbation when $\partial\vbar{U}$ coincides with large-amplitude regions of $\bsb{\psi}_{i}$.

        The analysis described in this section holds exactly for infinitesimally small perturbations in $\vbar{U}$. In practice, the predictions described herein are first-order approximations of the true perturbations. In the results that follow, it will be shown that the first-order approximations can accurately predict the perturbed quantities from resolvent analysis. To reflect the finite perturbation $\Delta \vbar{U}$ to $\vbar{U}$, the $\partial\vbar{U}$'s will be replaced with $\Delta\vbar{U}$'s without loss of generality though keeping in mind that the expressions in Equations \ref{Eq_SimpSensitivitySigma}--\ref{Eq_SimpSensitivityPhi} are only first-order in $\Delta \vbar{U}$. In cases where $\vbar{U}$ is parameterized by parameters $\vb{a}$, $\Delta \vbar{U}$ can either be computed as a difference as $\Delta\vbar{U} = \vbar{U}(\vb{x}_{n};\vb{a} + \Delta\vb{a}) - \vbar{U}(\vb{x}_{n};\vb{a})$ or as a first-order approximation with $\Delta\vbar{U} \approx \pdv*{\vbar{U}(\vb{x}_{n};\vb{a})}{\vb{a}}\Delta\vb{a}$.

        For a TBL, $\oline{U}\gg\oline{V}$ and $\oline{U}_{y}\gg\oline{U}_{x}\gg\oline{V}_{x}$ and $\mb{W}_{f} = \mb{B}$, equation \ref{Eq_SimpSensitivitySigma} can be approximated as 
            \begin{equation}
                \Delta\sigma_{i} \approx -\sigma_{i}^{2}\Real\qty(  \int \bsb{\phi}_{i}^{*}\mb{B}\Delta\oline{U}\hat{\partial}_{x}\bsb{\psi}_{i} + \phi^{*}_{u,i}\Delta\oline{U}_{y}\psi_{v,i}  d\vb{x}_{h} ),
            \end{equation}
        which is exact for the local analysis. Here, $\what{\partial}_{x} = ik_{x}$ for the local approach and $\what{\partial}_{x} = \partial_{x}$ in the biglobal approach. Thus, the near-wall error primarily affects the streamwise advection and the mean shear. Furthermore, $\Delta\sigma_{i}$ is non-negligible if $\bsb{\phi}_{i}$ and $\bsb{\psi}_{i}$ are non-zero where $\Delta\oline{U}$ and $\Delta\oline{U}_{y}$ is present.

    \subsection{Mean flow fields that model near-wall uncertainty and PIV set up} \label{Sec_NearWallFits}
        We first model common sources of experimental error in near-wall measurements through uncertainty in the absolute position of the wall-normal measurements and uncertainty in $\tau_{w}$. We begin with reference data from the direct numerical simulation (DNS) of the ZPG TBL of \citet{schlatter2010assessment} using their mean flow data at $Re_{\tau} = 1270$, which will be here denoted as $\oline{U}_{0}(y)$. In what follows, we model uncertainty in the near-wall data by removing the reference data for $y^{+}<150$ and introducing artificial errors when we apply a near-wall fit to our retained data. We will consider these erroneously fitted means as perturbations to the mean. The viscous scaled quantities denoted by $+$ are normalized by the unperturbed friction velocity, $u_{\tau,0}$, and viscous length scale, $\ell_{\nu,0}$ used to compute the reference $Re_\tau$.

        To model the uncertainty in the wall-normal position, we offset the vertical position of the retained data by $\Delta y^+ = 36$ as $\oline{U}(y;\pm\Delta y) = \oline{U}(y^{+} \pm 36)$. We then fit the lowest retained point to the reference DNS data to determine $u_{\tau}$ and obtain a near-wall mean velocity estimate. A positive offset produces a $\Delta u_{\tau}/u_{\tau,0} = -0.035$ and a negative offset produces a $\Delta u_{\tau}/u_{\tau,0} = 0.026$.

        To model the effect of uncertainty in $u_{\tau}$, we fit the lowest point in the reference data to a perturbed wall function. Here, we take our perturbed wall function to be $(1 + \gamma)\oline{U}_{0}((1 + \gamma)y)$, which has the effect of producing an erroneous $u_{\tau}$. Though this is a poor choice of a wall function, it is representative of the effect that different wall functions can have on the mean flow measurements. We choose a value of $\gamma = 0.225$ and $-0.225$, producing a $\Delta u_{\tau}/u_{\tau,0} = 0.21$ and $-0.14$, respectively. Compared to the vertical offset, the change to the wall function has an expected substantial effect on the shear stress, and consequently, the near-wall region. Hereafter, we will parameterize the change in $\oline{U}$ from $\gamma$ by $\Delta u_{\tau}$ as $\oline{U}(y;\Delta u_{\tau})$ to highlight the effect of $\gamma$ on the near-wall reconstruction.

            \begin{figure}
                \centering
                \includegraphics[width=0.95\linewidth]{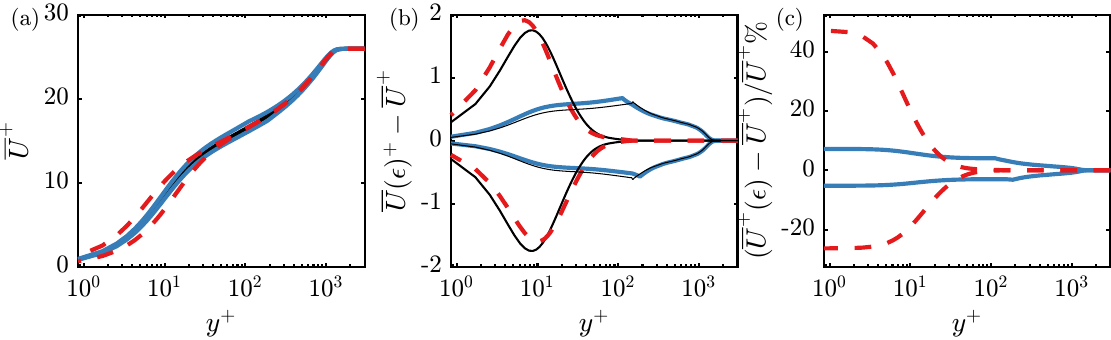}
                \caption{(a) $\oline{U}^{+}$ vs $y^{+}$ for the unperturbed flow $\oline{U}_{0}$ (black), perturbation to $u_{\tau}$ (red), and offset in $y$ (blue). (b) difference in perturbed flow fields and unperturbed flows with thin black lines representing $\pdv*{\oline{U}}{a}\Delta a$. (c) percent difference in perturbed flow fields and unperturbed flows. }
                \label{fig_MeanFlowFields}
            \end{figure}

        The perturbed and unperturbed mean flow fields are plotted in figure \ref{fig_MeanFlowFields}(a), along with measures of $\Delta\oline{U}$ in figures \ref{fig_MeanFlowFields}(b,c). For $\oline{U}(y;\pm\Delta u_{\tau})$, the error is concentrated in the near-wall region. The error in $\oline{U}(y;\pm\Delta y)$ is distributed throughout the boundary layer, though its largest percent error is again in the near-wall region. Additionally, we include a first order approximation of the error in the mean flow field via $\pdv*{\oline{U}(y;a)}{a} \Delta a$, where $a$ is either $\Delta y$ or $\Delta u_{\tau}$ in figure \ref{fig_MeanFlowFields}(b). The difference between $\Delta \oline{U}$ and $\pdv*{\oline{U}(y;a)}{a} \Delta a$ highlight that this is not a first order perturbation in $\oline{U}$. In fact, the errors are near $10\%$ for $\oline{U}(y;\pm\Delta y)$ and up to $40\%$ for $\oline{U}(y;\pm\Delta u_{\tau})$ in figure \ref{fig_MeanFlowFields}(c)

       For the non-parallel case, we employ an experimental PIV dataset at $Re_{\tau} = 1230$ to illustrate a realistic near-wall TBL. The experimental parameters and conditions are those that can be found in a standard TBL experiment where resources may not be available to achieve fully spatially and temporally resolved near-wall data. The experiment was conducted in a closed-return wind tunnel equipped with an optical-access test section. A Photonics DM-50-527 frequency-doubled Nd:YLF pulsed laser ($\lambda = 532~\text{nm}$) generated a light sheet of approximately $1~\text{mm}$ thickness, introduced from the top of the test section to define the $x$–$y$ measurement plane. The flow was seeded with \textit{di-ethylhexyl sebacate} (DEHS) tracer particles, with a nominal diameter of $1~\mu\text{m}$. Time-resolved PIV measurements were acquired using a Phantom V2012 high-speed camera synchronized with the laser system. Image pairs were recorded at 10~kHz for a total acquisition duration of approximately 5~s. The PIV images were processed using \textsc{Davis} software (LaVision GmbH). Velocity vectors were computed with a multi-pass cross-correlation scheme, employing final interrogation windows of $32 \times 32$ pixels with 50\% overlap. This yielded a vector spacing corresponding to 0.3 mm. The Reynolds number based on the momentum thickness was later evaluated to ensure consistency with canonical turbulent boundary layer datasets. The measured dataset shows good agreement with both canonical experiments and DNS simulations of the mean flow field. The wall-normal point closest to the wall in the PIV data is $2.448$ (mm) from the wall or $145\ell_{\nu}$.

       In this case, we need to perform a near-wall fit to obtain an accurate estimate of the near-wall flow field. Here, we explore two options. The first fits the PIV data to the reference DNS data, $\oline{U}_{0}(y)$, at each streamwise location to obtain a an estimate of $u_{\tau}(x)$ and subsequently, an estimate of $\oline{U}(x,y)$ throughout the full domain and boundary layer. This mean flow field is denoted as $\oline{U}_{1}$. For the second option, we use the \citet{van1956turbulent} mixing length, $\ell_{m}^{+}(y^{+}) = \kappa y^{+} \qty(1 - \exp(-y^{+}/A^{+}))$, to obtain the wall-function,
            \begin{equation}
                f_{w}^{+}(y^{+}) = \int_{0}^{y^+} \frac{2 dy'}{1 + \sqrt{\qty[1 + 4 \ell_{m}^{+}(y')^{2}]}},
            \end{equation}
        with $\kappa = 0.4$ and $A^{+} = 26$~\citep{pope2000turbulent}. At each streamwise location, we then fit $f_{w}^{+}(y^{+})$ to the PIV data to obtain a new estimate of $u_{\tau}(x)$ and the mean flow field, here denoted as $\oline{U}_{2}(x,y)$. Finally, since the streamwise domain of the PIV data is limited, we artificially extend the mean fields by setting $\vbar{U}$ upstream and downstream of the PIV domain as streamwise constant, which can be justified by the relatively slow boundary layer growth. We then extract a streamwise domain that is extended two $\delta_{99}$ upstream and two $\delta_{99}$ downstream of the PIV measurement plane to artificially extend our computational domain. While this creates artificial effects, the differences between $\vbar{U}_{1}$ and $\vbar{U}_{2}$ are due to the near-wall fits. The difference in $\oline{U}_{1}$ and $\oline{U}_{2}$ is around $3\%$ in the near-wall region, which will be discussed in more detail in section \ref{Sec_Res2D}.

\section{Results}\label{Sec_Res}
    
    \subsection{Local analysis}\label{Sec_Res1D}

            \begin{figure}
                \centering
                \includegraphics[width=0.99995\linewidth]{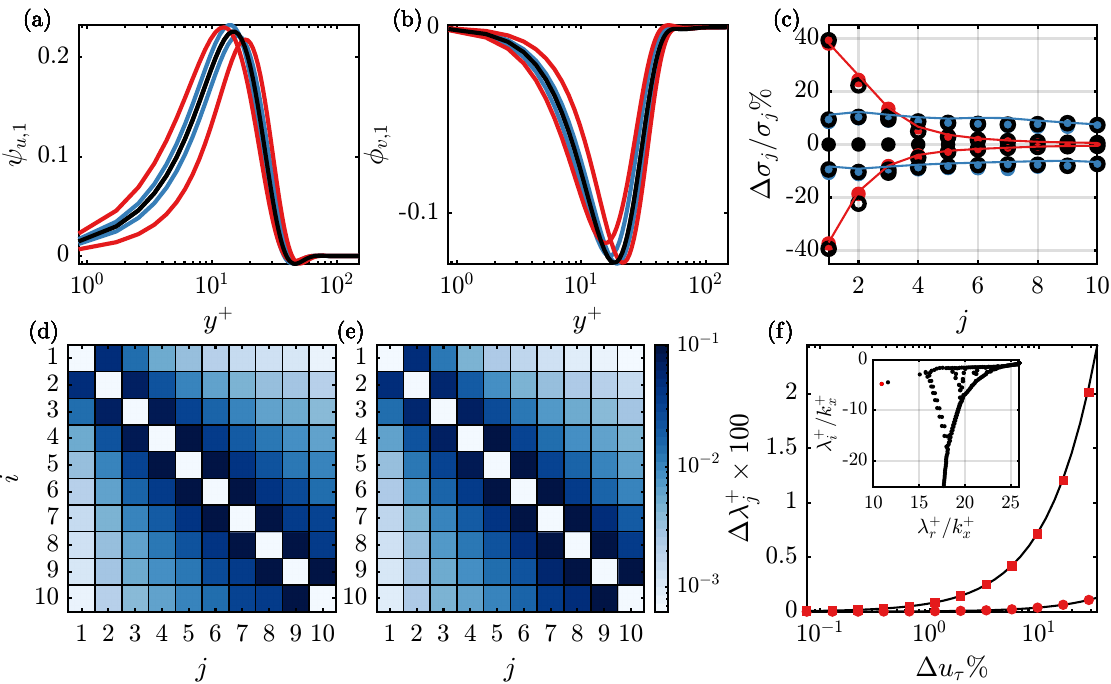}
                \caption{Comparison of 1D $\Real(\psi_{u,1})$ (a) $\Real(\phi_{v,1})$ (b), and the percent difference of $\sigma_{j}$ computed from the perturbed flows relative to the unperturbed $\oline{U}$ (c) computed from unperturbed and perturbed $\oline{U}$. The colors are the same as in Figure \ref{fig_MeanFlowFields}. The predicted perturbed $\sigma_{j}$, $\sigma_{j} + \Delta\sigma_{j}$, are plotted in black circles in (c). Comparison of $\ip{\bsb{\psi}_{j}}{\mb{W}_{r}\Delta\bsb{\psi}_{i}}$ using the prediction (d) and actual difference (e) for a perturbation in $\Delta y$. Comparison of $\Delta \lambda_{j}$ (red squares) and $\Delta \sigma_{1}/\sigma_{1}^{2}$ (red circles) against $\Delta u_{\tau}$ where $j$ is chosen as the $\Real(\lambda^{+}_{j}/k_{x}^{+})$ closest to $10$ in (f). The solid black lines are the predicted $\Delta \lambda_{j}$ and $\Delta \sigma_{1}/\sigma_{1}^{2}$. The resolvent analysis and stability analysis are computed with $k_{x}^{+} = 2\pi/1000$ and $k_{z}^{+} = 2\pi/100$ and the former uses $\omega^{+} = 10 k_{x}^{+}$.}
                \label{fig_1DModesComp}
            \end{figure}

        We first begin with a mode representative of the near-wall cycle with $k_{x}^{+} = 2\pi/1000$, $k_{z}^{+} = 2\pi/100$, and $\omega^{+} = 10k_{x}^{+}$ to motivate the uncertainty quantification. For each mean flow field, we compute $\bsb{\psi}_{i}(y,k_{x},k_{z},\omega;a)$, $\bsb{\phi}_{i}(y,k_{x},k_{z},\omega;a)$, and $\sigma_{i}(k_{x},k_{z},\omega;a)$ and use the notation $\Delta f = f(a)-f(0)$ to define the difference of the resolvent outputs in the perturbed mean relative to the unperturbed mean. Here, $a = \Delta y$ or $\Delta u_{\tau}$ as in section \ref{Sec_NearWallFits}. In figure \ref{fig_1DModesComp}(a,b), the effect of the different $\oline{U}$ on the dominant resolvent modes alters their amplitudes and peak locations by about $5$ viscous units, though they maintain their overall shape despite the significant change in the mean. In figure \ref{fig_1DModesComp}(c), the $\Delta \sigma_{i}/\sigma_{i}(k_{x},k_{z},\omega;0)$ are plotted, demonstrating $\mc{O}(10\%)$ changes in the singular values. Using equation \ref{Eq_SimpSensitivitySigma}, this perturbation in the singular values can be predicted using only the results from the unperturbed mean and knowledge of the perturbation to $\oline{U}$. In a similar vein, figures \ref{fig_1DModesComp}(d,e) compare $\ip{\bsb{\psi}_{j}}{\mb{W}_{r}\Delta\bsb{\psi}_{i}}$ using the prediction from equation \ref{Eq_SimpSensitivityPsi} and  the actual difference in the modes using $\oline{U}(y;\Delta y)$. This sensitivity analysis of the resolvent modes demonstrates that $\Delta \bsb{\psi}_{i}$ and $\Delta\bsb{\phi}_{i}$ will primarily have a large component in $\bsb{\psi}_{i+1}$ and $\bsb{\phi}_{i+1}$, respectively. Thus for a significant perturbation in $\oline{U}$, this analysis predicts the perturbations of $\sigma_{i}$, $\bsb{\psi}_{i}$, and $\bsb{\phi}_{i}$.

        We now compare the sensitivity of linear stability theory with $\sigma_{i}$. This involves the eigenvalues, $\lambda_{i} \in \mathbb{C}$, and eigenmodes, $\vb{v}_{i}$, of the generalized eigenvalue problem, $\vb{L}\vb{v}_{i} = \lambda_{i}\mb{B}\vb{v}_{i}$, as well as the adjoint eigenmodes, $\vb{w}_{i}$ to compute $\delta \lambda_{i} = \ip{\vb{w}_{i}}{\vb{B}\delta\vb{L}\vb{v}_{i}}/\ip{\vb{w}_{i}}{\vb{B}\vb{v}_{i}}$. We compare the sensitivity of $\lambda_{i}$ to $\sigma_{i}^{-1}$ as both quantities have units of frequency. Thus, we compare $\Delta\lambda_{i}$ and $\Delta\sigma_{1}^{-1} = -\Delta\sigma_{1}/\sigma_{1}^{2}$. Since $\lambda_{j}$ is parameterized by $k_{x}$ and $k_{z}$, to obtain a relevant comparison to $\sigma_{1}(k_{x},k_{z},\omega)$, we choose $j$ such that $\Real(\lambda^{+}_{j}/k_{x}^{+})$ is closest to $\omega^{+}/k_{x}^{+}$ so that the modes convect at the same velocity. The eigenspectra is shown in the inset of figure \ref{fig_1DModesComp}(f), with the identified mode shown in red. We note that this mode is not representative of the sensitivity of the stability analysis since it is far from the intersection of the three branches, the region most sensitive to perturbation~\citep{reddy1993pseudospectra,trefethen1999spectra}. In figure \ref{fig_1DModesComp}(f), $\Delta \lambda_{j}$ and $\abs{\Delta\sigma_{1}^{-1} }$ are compared for varying $\oline{U}(y;\Delta u_{\tau})$ and a wide range of $\Delta u_{\tau}$. The predicted perturbations match the computed perturbations in the two quantities and also highlight the increased sensitivity in the eigenvalues compared to the singular value.

        Now we highlight the spectral and wall-normal regions that these near-wall errors can affect when computing resolvent analysis. The critical layer, $y_{c}$ where $\oline{U}(y_{c}) = \omega/k_{x}$, is a a source of significant amplification in the resolvent operator which can in turn cause $\bsb{\phi}_{1}$ and $\bsb{\psi}_{1}$ to peak at $y_{c}$~\citep{mckeonsharma2010}. We use the critical layer as an estimate for the wall-normal position of the resolvent modes. The resolvent sweeps use $56$ log-spaced $\lambda_{x}^{+}$ between $10$ and $10^5$, $72$ log-spaced $\lambda_{z}^{+}$ between $10$ and $10^5$, fixed $\omega^{+}/k_{x}^{+}$ using $\oline{U}(y;0)$ and $\oline{U}(y;\Delta u_{\tau})$. Here, $\omega^{+}/k_{x}^{+} = 10$, $18$, and $23$, which correspond to $y_{c}^{+}=13$, $200$, and $750$, respectively.

            \begin{figure}
                \centering
                \includegraphics[width=0.95\linewidth]{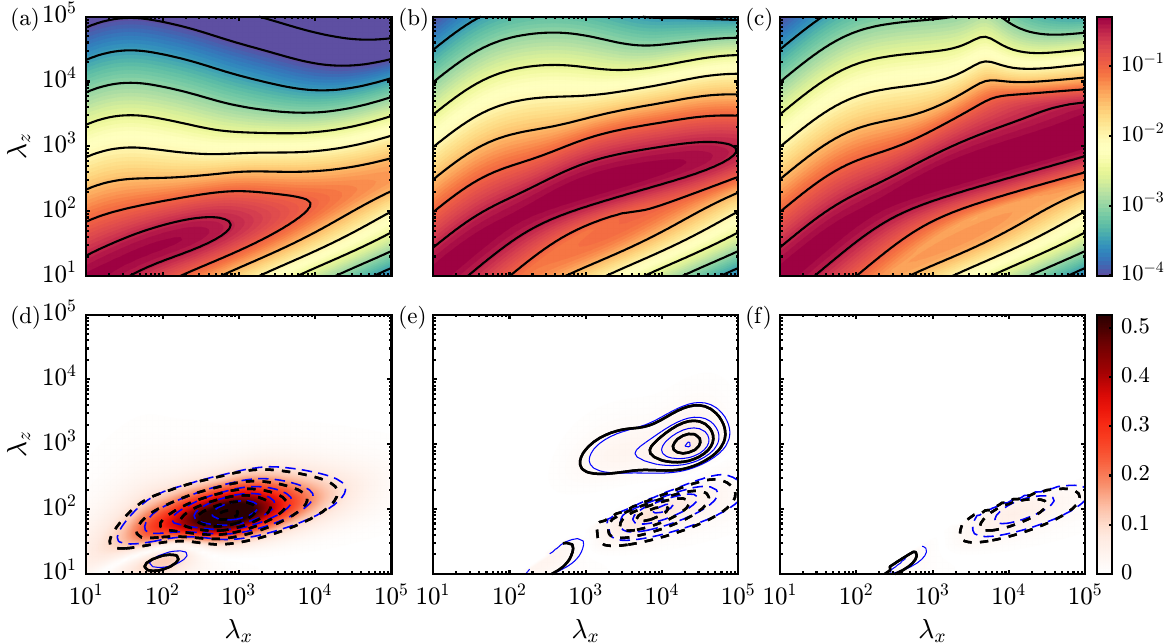}
                \caption{ Contours of $\sigma_{1}^{+}k_{x}^{+}k_{z}^{+}$ and $k_{x}^{+}k_{z}^{+}\partial\sigma_{1}^{+}/\partial u_{\tau}$ for fixed $\omega^{+}/k_{x}^{+} = 10$ ($y_{c}^{+} =13$) (a,d), $\omega^{+}/k_{x}^{+} = 18$ ($y_{c}^{+} = 200$) (b,e), and $\omega^{+}/k_{x}^{+} = 23$ ($y_{c}^{+} = 750$) (c,f). The colored contours plot $k_{x}^{+}k_{z}^{+}\abs{\partial\sigma_{1}^{+}/\partial u_{\tau}}$ while the solid contour lines are positive and dashed contours are negative in (d--f). The black contours are computed using the derivatives and the blue are computed using finite differences. The contour lines are in increments of $.1$ and $.02$ in (d) and (e,f), respectively. }
                \label{fig_Resolvent1DSweeps}
            \end{figure}

        In figure \ref{fig_Resolvent1DSweeps}(a--c) $k_{x}^{+}k_{z}^{+}\sigma^{+}_{1}$ is plotted for three different wavespeeds corresponding to different $y_{c}$. The large scale structures are more amplified for the $y_{c}$ further from the wall. As evidenced in figure \ref{fig_Resolvent1DSweeps}(d--f), $k_{x}^{+}k_{z}^{+}\partial\sigma_{1}^{+}/\partial u_{\tau}$ attains larger magnitudes for the $y_{c}$ closer to the wall since the resolvent modes peak in the region where $\Delta\oline{U}$ is largest. Additionally, the smaller $\lambda_{x}$ and $\lambda_{z}$ tend to have the largest sensitivity in $k_{x}^{+}k_{z}^{+}\sigma^{+}_{1}$, peaking around $(\lambda_{x}^{+},\lambda_{z}^{+}) = (1000,80)$, which is representative of scales present in the near-wall cycle~\citep{moarref2013model,hoyas2006scaling}. 
        For $y_{c}^{+}=200$ in figure \ref{fig_Resolvent1DSweeps}(e), the large-scale modes are sensitive to the near-wall perturbation since they have increased wall-normal extent, reaching down to the wall. For $y_{c}^{+} = 750$, the sensitivity has decreased substantially. Figures \ref{fig_Resolvent1DSweeps}(d--f) illustrate that the near-wall small-scale modes are most sensitive to the near-wall perturbation while the modes further from the wall are less influenced by the perturbation to the mean.

        In the local resolvent analysis, the resolvent modes are anchored to the critical layer making their wall-normal location predictable. Thus, modes with a critical layer in a region with large uncertainty in $\vbar{U}$ are expected to have the most sensitivity. In Section \ref{Sec_Implications}, we will address a practical example where error in $\oline{U}$ affects conclusions drawn from $\sigma_{i}$ between an APG and ZPG TBL.

    \subsection{Biglobal analysis}\label{Sec_Res2D}
        We apply the sensitivity analysis described in section \ref{Sec_MethSA} to a nonparallel ZPG TBL using PIV data that suffers from near-wall uncertainty in the mean flow. We define $\Delta \vbar{U} = \vbar{U}_{2} - \vbar{U}_{1}$ with the caveat that the true mean flow field, $\vbar{U}$ in the near-wall region is not measured.

            \begin{figure}
                \centering
                \includegraphics[width=0.95\linewidth]{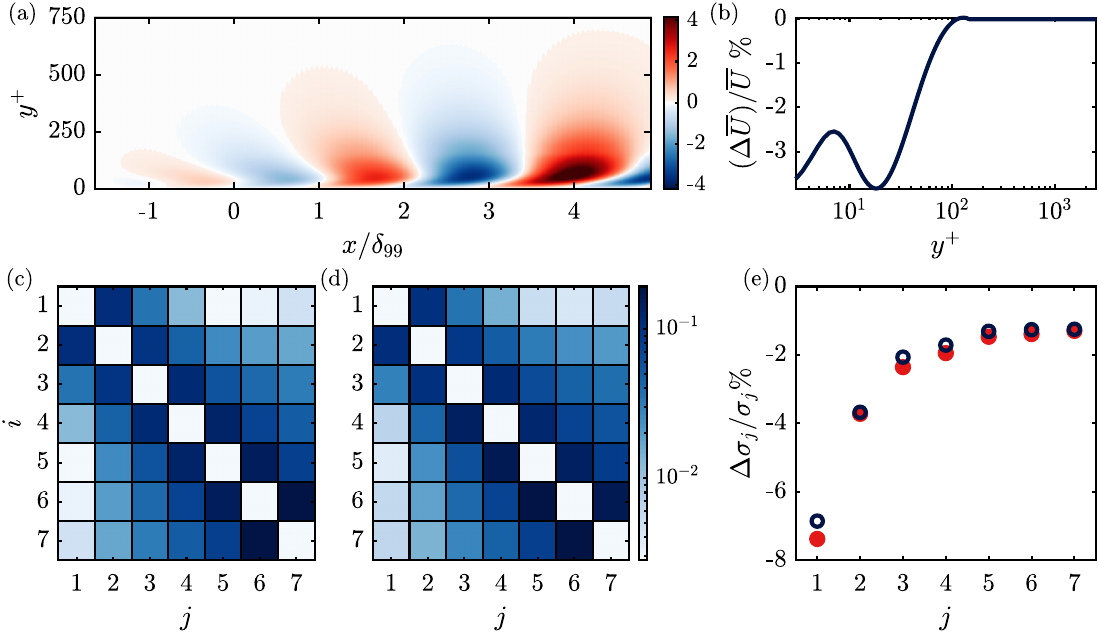}
                \caption{ Contour of $\Real(\psi_{u,1})$ (a). The percent difference in $\oline{U}$ by fitting to a DNS profile and the van Driest near-wall fit (b). Comparison of $\ip{\bsb{\psi}_{j}}{\mb{W}_{r}\Delta\bsb{\psi}_{i}}$ using the prediction (c) and actual difference (d) when using the two different near-wall fits. The percent difference in $\Delta \sigma_{j}$ computed from the prediction (black circles) and the true difference (red circles). }
                \label{fig_Resolvent2DIndividual}
            \end{figure}

        First, we illustrate the streamwise component of a representative response mode computed with $k_{z}\delta_{99} = 2\pi$ ($k_{z}^{+} = 2\pi/1240$) and $\omega U_{\infty}/\delta_{99} = 1.45$ ($\omega^{+} = 200$) representative of large-scale streaks in figure \ref{fig_Resolvent2DIndividual}(a) computed using $\vbar{U}_{1}$. The difference in the near-wall fits used is around $3\%$ as shown in figure \ref{fig_Resolvent2DIndividual}(b). These differences are much smaller than those used in Section \ref{Sec_Res1D} and are representative of expected differences in near-wall estimates used in practice. In figures \ref{fig_Resolvent2DIndividual}(c,d), the projection of $\Delta\bsb{\psi}_{i}$ onto $\bsb{\psi}_{j}$ is compared using the true difference and the prediction from equation \ref{Eq_SimpSensitivityPsi} showing excellent agreement for $i\ne j$. The $\Delta\sigma_{i}$ is also shown to be well predicted using equation \ref{Eq_SimpSensitivitySigma} for this nonparallel case in figure \ref{fig_Resolvent2DIndividual}(e). Despite the small $3\%$ difference in $\vbar{U}$ constrained near the wall, $\Delta\sigma_{1}/\sigma_1 \approx 8\%$ for this large-scale mode. Nonetheless, even for a $\Delta\vbar{U}(x,y)$ distributed across the entire domain, the error estimates can reasonably predict differences in the nonparallel resolvent modes.

            \begin{figure}
                \centering
                \includegraphics[width=0.9995\linewidth]{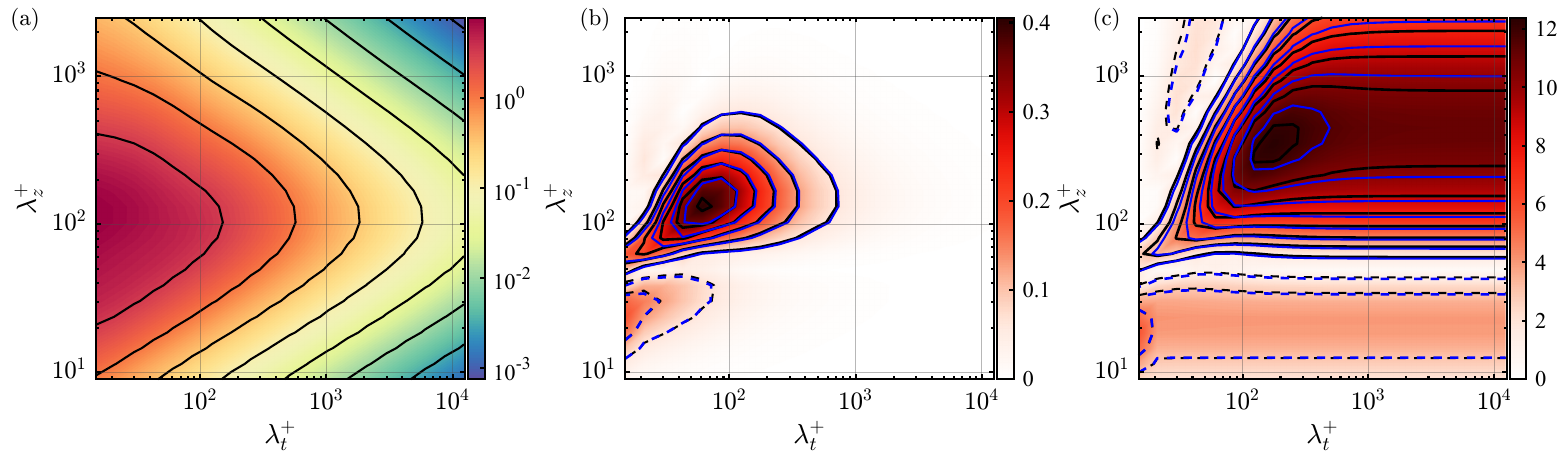}
                \caption{ Contours of $k_{z}^{+}\omega^{+}\sigma_{1}^{+}$ (a), $k_{z}^{+}\omega^{+}\Delta\sigma_{1}^{+}$ (b) and $\Delta\sigma_{1}/\sigma_{1} (\%)$ (c) for the PIV data with different near-wall fits. The solid and dashed contours correspond to positive and negative values spaced at increments of $.065$ in (b) and $1.5$ in (c). The blue contours are computed using finite differences. }
                \label{fig_Resolvent2DSweeps}
            \end{figure}

        Now we compute a sweep using $24$ logarithmically spaced $k_{z}^{+}$ in $\qty[2\pi/10,2\pi/11000]$ and $20$ logarithmically spaced $\omega^{+}$ in $\qty[2\pi/15,2\pi/12500]$ to identify the most sensitive $k_{z}$ and $\omega$. The premultiplied $\sigma_{1}$ are plotted in figure \ref{fig_Resolvent2DSweeps}(a), highlighting the increased amplification for $\lambda_{z}^{+}\approx 100$. The error is computed using $k_{z}^{+}\omega^{+}\Delta\sigma_{1}^{+}$ in figure \ref{fig_Resolvent2DSweeps}(b). The most sensitive scales are $\lambda_{z}^{+} \approx 150$ and $\lambda_{t}^{+}\approx 60$, which are length and time scales characteristic of the near-wall cycle. Since these scales are closest to the wall and scale with $u_{\tau}$~\citep{moarref2013model,moarref2014low_order_represenanation,gomez2024linear}, it is expected that these should be sensitive to the change in near-wall fit. The percent error is presented in figure \ref{fig_Resolvent2DSweeps}(c), which highlight a peak for $\lambda_{z}^{+}\approx 400$ and $\lambda_{t}^{+}\approx 150$. Unlike the local approach, $\Delta\sigma_{1}/\sigma_{1}$ is much larger for the larger scales in the biglobal approach. In the biglobal approach, this occurs because there is no explicit critical-layer amplification, thus modes are amplified by the non-normality which largely stems from the mean shear localized near the wall. Thus large scale modes, like the one shown in figure \ref{fig_Resolvent2DIndividual}(a), are localized near the wall and can also be influenced by the near-wall fit. $\Delta\sigma_{1}/\sigma_{1}$ begins to decay for the largest $\lambda_{z}$, suggesting that the largest scales can eventually lift away from the near-wall influence. In figures \ref{fig_Resolvent2DSweeps}(b,c), the predictions of $\Delta\sigma_{1}$ from equation \ref{Eq_SimpSensitivitySigma} agree with the $\Delta\sigma_{1}$ computed from true difference. 

    \subsection{Implications for Resolvent Analysis and Suggestions for Experimental Design} \label{Sec_Implications}

            \begin{figure}
                \centering
                \includegraphics[width=0.99\linewidth]{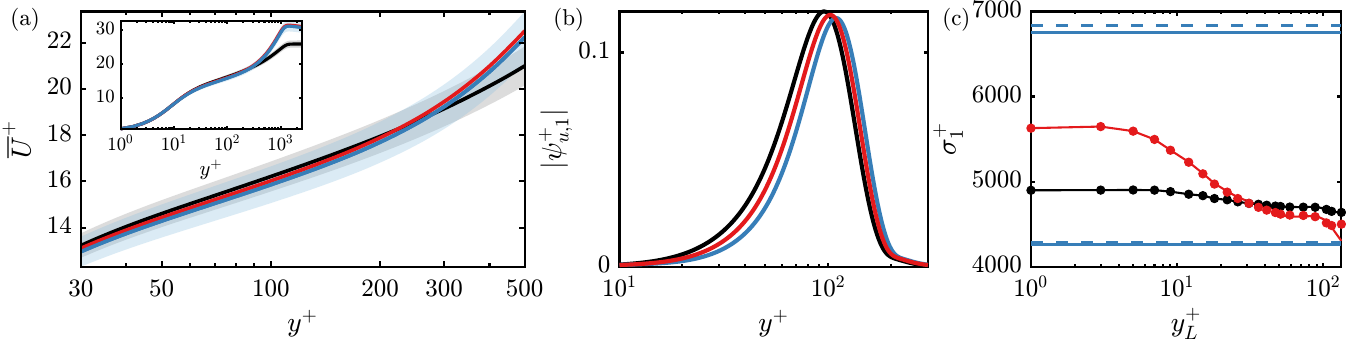}
                \caption{Comparison of $\oline{U}^{+}$ near the log-layer of a ZPG (black), APG (blue), and a profile fitted to the APG with $y_{L}^{+}=12.5$ (red) (a). The grey and blue shaded regions indicate a $4\%$ uncertainty in $\oline{U}$ and the inset shows the full mean profiles. The $|\psi_{u,1}|$ are computed using the $\oline{U}$ in (a) with the same color codes (b). The computed $\sigma_{1}(y_{L})$ (circles) and estimated $\sigma_{1}(0) + \Delta\sigma_{1}(y_L)$ (lines) using the fitted APG (red) and ZPG (black) $\oline{U}$ with varying $y_{L}^{+}$ (c). The solid blue lines plot $\sigma_{1}^{+}$ computed using $1.04\oline{U}$ and $0.96\oline{U}$ of the APG as lower- and upper-bound estimates of $\sigma_{1}^{+}$. The dashed blue lines are use $\sigma_{1}^{+} + \Delta\sigma_{1}^{+}$ as estimates. } 
                \label{fig_Resolvent1DExampleWarning}
            \end{figure}

        Now that we have shown that the near-wall sensitivity can be estimated and quantified, we illustrate a cautionary example where the mean flow uncertainty can lead to incorrect conclusions. We illustrate this by using the wall-resolved large eddy simulation data of an APG TBL of \citet{pozuelo2022adverse} and a ZPG TBL of \citet{eitel2014simulationZPG} at $Re_{\tau} = 1250$. While the APG primarily affects the wake in $\oline{U}$, it is known to reduce $\oline{U}^{+}$ in the log-layer relative to a ZPG. This is shown in figure \ref{fig_Resolvent1DExampleWarning}(a) where this difference is shown to fall within $4\%$ of the data. $\oline{U}$ uncertainties of approximately 1–5\% are typical for laboratory PIV~\citep{benedict1996towards} and $5\%$ errors in $u_{\tau}$ are common\citep{patel1965calibration,musker1979explicit,fernholz1996new,nagib2007approach,dixit2009determination}. We then introduce near-wall uncertainty by omitting the data below $y_{L}^{+}$ to mimic the near-wall resolution in experiment. We then fit the retained data to the reference DNS data as was done for the PIV data. In this section, the viscous-scaled quantities are normalized using the $u_{\tau}$ from the fitting procedure rather than the unperturbed $u_{\tau}$. In figure \ref{fig_Resolvent1DExampleWarning}(a), fitting the APG data with $y_{L}^{+}$ produces a $\oline{U}^{+}$ that falls between the ZPG and APG in the log-layer.

        We now use local analysis with the same computational grid described in section \ref{Sec_MethRA} and a mode representative of the log-layer with $k_{x}^{+} = 2\pi/1600$ $k_{z}^{+} = 2\pi/500$, and $\omega^{+} = 2\pi/100$. The representative modes are plotted in figure \ref{fig_Resolvent1DExampleWarning}(b), showing that the fitted mean profile with $y_{L}^{+} = 12.5$ falls between the APG and ZPG modes. We then vary $y_{L}^{+}$ for both the ZPG and APG data and plot $\sigma_{1}$ in figure \ref{fig_Resolvent1DExampleWarning}(c). From the unperturbed $\oline{U}$, one would conclude that the APG produces more amplification than the ZPG. However, when the data is omitted as low as $y_{L}^{+} = 30$, these conclusions are reversed. We also highlight that this change in the $\sigma_{1}$ can be predicted using equation \ref{Eq_SimpSensitivitySigma}. While it is true that fitting APG data to a ZPG reference data may produce erroneous profiles, other sources of error in experimental configurations can introduce uncertainty in the $\oline{U}^{+}$ profile. We illustrate a crude upper- and lower-bound estimate by using $1.04\oline{U}^{+}$ and $0.96\oline{U}^{+}$ in the calculation of $\sigma_{1}$ for the APG data as a surrogate for a $4\%$ error in the mean in figure \ref{fig_Resolvent1DExampleWarning}(c). The change in $\sigma_{1}$ with $y_{L}$ falls within this error bound and we note that this error bound can also be estimated \textit{a priori} using equation \ref{Eq_SimpSensitivitySigma} as well.

        In practical applications, resolvent analysis can also be used to create a model for the velocity fluctuations from the resolvent mode basis~\citep{moarref2013model,moarref2014low_order_represenanation,towne2018spectral}. The Fourier modes can be written as $\vhat{q} = \sum_{i}\sigma_{i}\chi_{i}\bsb{\psi}_{i}$ where $\chi_{i}$ are coefficients. The simplest approach to find $\chi_{i}$ involves exploiting the orthonormality of $\bsb{\psi}_{i}$ such that $\chi_{i}= \sigma_{i}^{-1}\ip{\vhat{q}}{\mb{W}_{r}\bsb{\psi}_{i}}$. For the systems described herein, $\chi_{i}$ has sensitivity
            \begin{equation}
                 \partial\chi_{i} = -\frac{\partial \sigma_{i}}{\sigma_{i}}\chi_{i}  + \frac{1}{\sigma_{i}}\sum_{j}\chi_{j}\sigma_{j}\ip{\bsb{\psi}_{i}}{\mb{W}_{r}\partial \bsb{\psi}_{i}} + \ip{\partial \vhat{q}}{\mb{W}_{r}\bsb{\psi}_{i}}, 
            \end{equation}
        where we here assume that $\vhat{q}$ can also be influenced by measurement or model error. Ignoring $\partial\vhat{q}$, the error in $\chi_{i}$ can be approximated using equations \ref{Eq_SimpSensitivitySigma} and \ref{Eq_SimpSensitivityPsi}. For realistic applications, like the PIV example in section \ref{Sec_Res2D} and the APG TBL example herein, both $\partial\sigma_{i}/\sigma_{i}$ and $\ip{\bsb{\psi}_{i}}{\mb{W}_{r}\partial \bsb{\psi}_{i}}$ can be $\mc{O}(10^{-1})$ highlighting how the true expansion coefficients can change due to the mean flow error that affects the resolvent basis. This can affect the number of resolvent modes one uses to represent the velocity fluctuations or require recalibration of the fitting or modeling procedure. While the flow may be reconstructed using the altered basis, any conclusions drawn from analyzing $\chi_{i}$ must keep in mind potential mean flow error similar to any conclusions drawn based on $\sigma_{i}$.

        We have highlighted how sources of experimental near-wall uncertainty can alter conclusions drawn from local resolvent analysis of log-layer modes if one is not careful with the fitting procedure in the near-wall or from reported measurement uncertainties~\citep{benedict1996towards}. Because of the critical layer amplification, issues in the mean flow field measurement can potentially be avoided. That is, if one knows that there is poor resolution within a certain wall-normal range, the $y_{c}$ can be chosen to be sufficiently far away from that source of uncertainty. It was shown in figure \ref{fig_Resolvent1DSweeps} that even for a large $\Delta\oline{U}$, moving sufficiently far away from the near-wall region produces almost negligible error in $\sigma_{1}$. For a local resolvent analysis, as long as the convective velocities correspond to a $y_{c}$ that falls in a region that is well measured, one can expect the analysis will not be prone to significant errors. For near-wall, small-scale modes, high near-wall resolution is required to avoid measurement errors. For biglobal resolvent analysis, the observations in figure \ref{fig_Resolvent2DSweeps}(c) illustrate that this near-wall uncertainty affects even the large-scale modes. Indeed, the scaling in \citet{gomez2025linearAPGTBL} showed that $\oline{U}_{y}$ from as low as $y^{+} = 40$ contributes to the large-scale amplification of biglobal modes in quantifying upstream effects. Thus, we suggest that for biglobal analysis, high-spatial resolution in the near-wall region is important to definitively differentiate the effects in the TBL from the true differences in the mean flow field and the experimental errors.

        Resolvent analysis is sensitive to the fidelity of the measured mean flow and to the spatial resolution of velocity gradients; in practice, expanding spatial coverage and improving gradient accuracy have a larger impact on resolvent predictions than increasing temporal sampling. Experimental design should therefore prioritize accurate, well-resolved mean fields. Means must be rigorously converged: obtain records long enough that doubling the record length changes the estimated mean by less than about $0.5$--$1\%$ targeting on the order of 10000 eddy turnover times at each location. Spatial resolution should prioritize resolving the near-wall gradients for both local and biglobal analyses. We have demonstrated that resolvent predictions are especially sensitive to near-wall velocity and gradient errors: sub-millimetric uncertainty in the wall position or $y^{+}$ can materially alter inferred shear and distort the operator. Invest in near-wall calibration, optical alignment, and probe positioning, and apply any smoothing or regularization to mean profiles only after verifying that it preserves physically relevant gradients.

        In cases where high-spatial resolution is not possible, then we suggest that the sensitivity analysis presented herein can be used as an error bound at a negligible additional cost. This can help reason where differences due to the change in the mean flow field are definitive or a source of experimental uncertainty. Furthermore, in cases where the near-wall mean is uncertain, we suggest normalization with outer variables where there is increased spatial resolution to avoid contamination from an ill-measured $u_{\tau}$ in determining the scales of interest.

\section{Conclusions}\label{Sec_Conc}
    We have developed a sensitivity analysis for resolvent analysis which can compute sensitivity of the resolvent modes and singular values with little additional computational overheard by using only matrix multiplications via $\Delta\sigma_{i} = -\sigma_{i}^{2}\Real(\ip{\bsb{\phi}_{i}}{\mb{W}_{f}\Delta \mb{L} \bsb{\psi}_{i}})$. This sensitivity analysis is then applied to near-wall uncertainty stemming from either measurement error or near-wall fits. This methodology predicts the discrepancies in resolvent analysis when using a DNS mean profile under the parallel flow assumption with large perturbations representative of true experimental error introduced into the mean profiles. We then apply this framework to biglobal resolvent analysis using PIV mean flow data with differing near-wall fits showing excellent agreement with the true differences.

    Because of the critical layer amplification, the parallel flow analysis provides the opportunity to probe the sensitivity based on the wall-normal coordinate by specifying the wavespeed. This demonstrated that, as expected, the near-wall modes are most sensitive to the near-wall perturbations. Further away from the wall, the sensitivity decreased substantially. The biglobal approach found large sensitivity for small scales representative of the near-wall cycle, which take advantage of the amplification from the mean shear. The large-scales are also amplified by the mean shear, making them susceptible to the near-wall uncertainty. These issues can also have implications for flow reconstruction where uncertainty in the expansion coefficients can also be tracked and related to uncertainty in the resolvent gains and modes.

    We also highlight an example where a poor near-wall fit leads to the wrong conclusions regarding the amplification of an APG and ZPG. Using this fit, the error in $\oline{U}^{+}$ was within $4\%$ of the true data and produced a change in $\sigma_{1}$ of around $20\%$. In the biglobal approach, the differences in $\oline{U}$ were around $3\%$ while the differences in $\sigma_{1}$ were up to $12\%$. This analysis should caution researchers using resolvent analysis with poorly resolved near-wall means like those measured in experiments or simulated using WMLES. Nonetheless, this analysis provides a method to estimate the expected error in resolvent analysis by estimating the mean flow uncertainty with little additional overhead.

\section*{Acknowledgment}
    S.R.G and T.J. gratefully acknowledge the support of ONR to CTR under grant N000142312833 and thank Professor Beverley McKeon and Professor Joe Klewicki for fruitful discussions during the preparation of this paper. 

\begin{appendix}
    \section{Uncertainty Quantification of Resolvent Analysis for General Flows}\label{App}
         For a general flow, $\mb{B}$ is now an identity operator in equation \ref{Meth_eq_FTNSE}. The LNSE and weight operators depend on $\vbar{Q}$ and parameters $\vb{a}$ as $\mb{W}_{r}(\vbar{Q},\vb{a})$ and $\mb{W}_{f}(\vbar{Q},\vb{a})$. This parameterization is especially relevant for compressible flows that often use \citet{chu1965energy}'s energy norm. Following similar steps as section \ref{Sec_MethSA}, except with nonzero $\partial\mb{W}_{r}$ and $\partial\mb{W}_{f}$, the perturbations of $\sigma_{i}$, $\bsb{\psi}_{i}$, and $\bsb{\phi}_{i}$ are now
            \begin{align}
                  \partial\sigma_{i}& = -\sigma_{i}^{2}\Real(\ip{\bsb{\phi}_{i}}{\mb{W}_{f}\partial \mb{A} \bsb{\psi}_{i}})
                  + \frac{\sigma_{i}}{2}\ip{\bsb{\psi}_{i}}{\partial \mb{W}_{r}\bsb{\psi}_{i}}
                  - \frac{\sigma_{i}}{2}\ip{\bsb{\phi}_{i}}{\partial \mb{W}_{f}\bsb{\phi}_{i}}, 
                  \label{Eq_FullSensitivitySigma} \\
                  \begin{split}
                        \ip{\bsb{\psi}_{j}}{\mb{W}_{r}\partial\bsb{\psi}_{i}}
                        &= \frac{1}{\sigma_{i}^{2}-\sigma_{j}^{2}}\bigg[
                            -\sigma_{i}^{2}\sigma_{j}\ip{\bsb{\phi}_{j}}{\mb{W}_{f}\partial\mb{A}\bsb{\psi}_{i}} 
                            -\sigma_{j}^{2}\sigma_{i}\ip{\partial\mb{A}\bsb{\psi}_{j}}{\mb{W}_{f}\bsb{\phi}_{i}} \\
                        &\quad\quad\quad
                           - \sigma_{j}\sigma_{i}\ip{\bsb{\phi}_{j}}{\partial\mb{W}_{f}\bsb{\phi}_{i}} + \sigma_{j}^{2}\ip{\bsb{\psi}_{j}}{\partial\mb{W}_{r}\bsb{\psi}_{i}}
                          \bigg].
                  \end{split}
                  \label{Eq_FullSensitivityPsi} \\
                  \begin{split}
                        \ip{\bsb{\phi}_{j}}{\mb{W}_{f}\partial\bsb{\phi}_{i}} &= \frac{1}{\sigma_{i}^{2}-\sigma_{j}^{2}}\bigg[-\sigma_{i}^{2}\sigma_{j}\ip{\partial\mb{A}\bsb{\psi}_{j}}{\mb{W}_{f}\bsb{\phi}_{i}}-\sigma_{j}^{2}\sigma_{i}\ip{\bsb{\phi}_{j}}{\mb{W}_{f}\partial\mb{A}\bsb{\psi}_{i}} \\ 
                        &\quad\quad\quad - \sigma_{i}^{2}\ip{\bsb{\phi}_{j}}{\partial\mb{W}_{f}\bsb{\phi}_{i}} + \sigma_{j}\sigma_{i}\ip{\bsb{\psi}_{j}}{\partial\mb{W}_{r}\bsb{\psi}_{i}}\bigg]. 
                  \end{split}
                  \label{Eq_FullSensitivityPhi}
            \end{align}
        Once again, we note that these equations do not involve $\mb{H}$ or $\mb{H}^{\dagger}$, thus computing $\partial\sigma_{i}$, $\bsb{\psi}_{i}$, and $\bsb{\phi}_{i}$ involves inner products with $\partial\mb{A}$ and resolvent modes and gains that would otherwise be computed.

    \section{Example code implementation}
        \begin{lstlisting}[style=MatlabStyle, caption={MATLAB implementation of equation \ref{Eq_SimpSensitivitySigma}}]
function [dsigma] = mySquare(sigma,psi,phi,Wf,dLHS)
    % sigma   - computed resolvent gains (n by 1)
    % psi,phi - computed response modes (N by n), forcing modes (N by n)
    % Wf      - weight matrix (N by N)
    % dLHS    - perturbed LHS matrix (N by N) computed using finite differences or coded
    dsigma = -(sigma.^2).*real(phi'*(Wf*dLHS*psi));
end
        \end{lstlisting}
\end{appendix}
\bibliography{sn-bibliography}


\begin{thebibliography}{56}
\ifx \bisbn   \undefined \def \bisbn  #1{ISBN #1}\fi
\ifx \binits  \undefined \def \binits#1{#1}\fi
\ifx \bauthor  \undefined \def \bauthor#1{#1}\fi
\ifx \batitle  \undefined \def \batitle#1{#1}\fi
\ifx \bjtitle  \undefined \def \bjtitle#1{#1}\fi
\ifx \bvolume  \undefined \def \bvolume#1{\textbf{#1}}\fi
\ifx \byear  \undefined \def \byear#1{#1}\fi
\ifx \bissue  \undefined \def \bissue#1{#1}\fi
\ifx \bfpage  \undefined \def \bfpage#1{#1}\fi
\ifx \blpage  \undefined \def \blpage #1{#1}\fi
\ifx \burl  \undefined \def \burl#1{\textsf{#1}}\fi
\ifx \doiurl  \undefined \def \doiurl#1{\url{https://doi.org/#1}}\fi
\ifx \betal  \undefined \def \betal{\textit{et al.}}\fi
\ifx \binstitute  \undefined \def \binstitute#1{#1}\fi
\ifx \binstitutionaled  \undefined \def \binstitutionaled#1{#1}\fi
\ifx \bctitle  \undefined \def \bctitle#1{#1}\fi
\ifx \beditor  \undefined \def \beditor#1{#1}\fi
\ifx \bpublisher  \undefined \def \bpublisher#1{#1}\fi
\ifx \bbtitle  \undefined \def \bbtitle#1{#1}\fi
\ifx \bedition  \undefined \def \bedition#1{#1}\fi
\ifx \bseriesno  \undefined \def \bseriesno#1{#1}\fi
\ifx \blocation  \undefined \def \blocation#1{#1}\fi
\ifx \bsertitle  \undefined \def \bsertitle#1{#1}\fi
\ifx \bsnm \undefined \def \bsnm#1{#1}\fi
\ifx \bsuffix \undefined \def \bsuffix#1{#1}\fi
\ifx \bparticle \undefined \def \bparticle#1{#1}\fi
\ifx \barticle \undefined \def \barticle#1{#1}\fi
\bibcommenthead
\ifx \bconfdate \undefined \def \bconfdate #1{#1}\fi
\ifx \botherref \undefined \def \botherref #1{#1}\fi
\ifx \url \undefined \def \url#1{\textsf{#1}}\fi
\ifx \bchapter \undefined \def \bchapter#1{#1}\fi
\ifx \bbook \undefined \def \bbook#1{#1}\fi
\ifx \bcomment \undefined \def \bcomment#1{#1}\fi
\ifx \oauthor \undefined \def \oauthor#1{#1}\fi
\ifx \citeauthoryear \undefined \def \citeauthoryear#1{#1}\fi
\ifx \endbibitem  \undefined \def \endbibitem {}\fi
\ifx \bconflocation  \undefined \def \bconflocation#1{#1}\fi
\ifx \arxivurl  \undefined \def \arxivurl#1{\textsf{#1}}\fi
\csname PreBibitemsHook\endcsname

\bibitem[\protect\citeauthoryear{Coles}{1956}]{coles1956law}
\begin{barticle}
\bauthor{\bsnm{Coles}, \binits{D.}}:
\batitle{The law of the wake in the turbulent boundary layer}.
\bjtitle{Journal of Fluid Mechanics}
\bvolume{1}(\bissue{2}),
\bfpage{191}--\blpage{226}
(\byear{1956})
\end{barticle}
\endbibitem

\bibitem[\protect\citeauthoryear{Van~Driest}{1956}]{van1956turbulent}
\begin{barticle}
\bauthor{\bsnm{Van~Driest}, \binits{E.R.}}:
\batitle{On turbulent flow near a wall}.
\bjtitle{Journal of the aeronautical sciences}
\bvolume{23}(\bissue{11}),
\bfpage{1007}--\blpage{1011}
(\byear{1956})
\end{barticle}
\endbibitem

\bibitem[\protect\citeauthoryear{Schmid and Henningson}{2002}]{schmid2002stability}
\begin{bbook}
\bauthor{\bsnm{Schmid}, \binits{P.J.}},
\bauthor{\bsnm{Henningson}, \binits{D.S.}}:
\bbtitle{Stability and Transition in Shear Flows}
vol. \bseriesno{142}.
\bpublisher{Springer}, \blocation{???}
(\byear{2002})
\end{bbook}
\endbibitem

\bibitem[\protect\citeauthoryear{Trefethen et~al.}{1993}]{trefethen1993hydrodynamic}
\begin{barticle}
\bauthor{\bsnm{Trefethen}, \binits{L.N.}},
\bauthor{\bsnm{Trefethen}, \binits{A.E.}},
\bauthor{\bsnm{Reddy}, \binits{S.C.}},
\bauthor{\bsnm{Driscoll}, \binits{T.A.}}:
\batitle{Hydrodynamic stability without eigenvalues}.
\bjtitle{Science}
\bvolume{261}(\bissue{5121}),
\bfpage{578}--\blpage{584}
(\byear{1993})
\end{barticle}
\endbibitem

\bibitem[\protect\citeauthoryear{Reddy et~al.}{1993}]{reddy1993pseudospectra}
\begin{barticle}
\bauthor{\bsnm{Reddy}, \binits{S.C.}},
\bauthor{\bsnm{Schmid}, \binits{P.J.}},
\bauthor{\bsnm{Henningson}, \binits{D.S.}}:
\batitle{Pseudospectra of the orr--sommerfeld operator}.
\bjtitle{SIAM Journal on Applied Mathematics}
\bvolume{53}(\bissue{1}),
\bfpage{15}--\blpage{47}
(\byear{1993})
\end{barticle}
\endbibitem

\bibitem[\protect\citeauthoryear{Del~Alamo and Jimenez}{2006}]{del2006linear}
\begin{barticle}
\bauthor{\bsnm{Del~Alamo}, \binits{J.C.}},
\bauthor{\bsnm{Jimenez}, \binits{J.}}:
\batitle{Linear energy amplification in turbulent channels}.
\bjtitle{Journal of Fluid Mechanics}
\bvolume{559},
\bfpage{205}--\blpage{213}
(\byear{2006})
\end{barticle}
\endbibitem

\bibitem[\protect\citeauthoryear{Cossu et~al.}{2009}]{cossu2009optimal}
\begin{barticle}
\bauthor{\bsnm{Cossu}, \binits{C.}},
\bauthor{\bsnm{Pujals}, \binits{G.}},
\bauthor{\bsnm{Depardon}, \binits{S.}}:
\batitle{Optimal transient growth and very large--scale structures in turbulent boundary layers}.
\bjtitle{Journal of Fluid Mechanics}
\bvolume{619},
\bfpage{79}--\blpage{94}
(\byear{2009})
\end{barticle}
\endbibitem

\bibitem[\protect\citeauthoryear{Hwang and Cossu}{2010}]{hwang2010linear}
\begin{barticle}
\bauthor{\bsnm{Hwang}, \binits{Y.}},
\bauthor{\bsnm{Cossu}, \binits{C.}}:
\batitle{Linear non-normal energy amplification of harmonic and stochastic forcing in the turbulent channel flow}.
\bjtitle{Journal of Fluid Mechanics}
\bvolume{664},
\bfpage{51}--\blpage{73}
(\byear{2010})
\end{barticle}
\endbibitem

\bibitem[\protect\citeauthoryear{Reynolds and Hussain}{1972}]{reynolds1972mechanics}
\begin{barticle}
\bauthor{\bsnm{Reynolds}, \binits{W.}},
\bauthor{\bsnm{Hussain}, \binits{A.}}:
\batitle{The mechanics of an organized wave in turbulent shear flow. part 3. theoretical models and comparisons with experiments}.
\bjtitle{Journal of Fluid Mechanics}
\bvolume{54}(\bissue{2}),
\bfpage{263}--\blpage{288}
(\byear{1972})
\end{barticle}
\endbibitem

\bibitem[\protect\citeauthoryear{McKeon and Sharma}{2010}]{mckeonsharma2010}
\begin{barticle}
\bauthor{\bsnm{McKeon}, \binits{B.J.}},
\bauthor{\bsnm{Sharma}, \binits{A.S.}}:
\batitle{A critical-layer framework for turbulent pipe flow}.
\bjtitle{Journal of Fluid Mechanics}
\bvolume{658},
\bfpage{336}--\blpage{382}
(\byear{2010})
\end{barticle}
\endbibitem

\bibitem[\protect\citeauthoryear{Symon et~al.}{2018}]{symon2018non}
\begin{barticle}
\bauthor{\bsnm{Symon}, \binits{S.}},
\bauthor{\bsnm{Rosenberg}, \binits{K.T.}},
\bauthor{\bsnm{Dawson}, \binits{S.T.M.}},
\bauthor{\bsnm{McKeon}, \binits{B.J.}}:
\batitle{Non-normality and classification of amplification mechanisms in stability and resolvent analysis}.
\bjtitle{Physical Review Fluids}
\bvolume{3}(\bissue{5}),
\bfpage{053902}
(\byear{2018})
\end{barticle}
\endbibitem

\bibitem[\protect\citeauthoryear{Moarref et~al.}{2013}]{moarref2013model}
\begin{barticle}
\bauthor{\bsnm{Moarref}, \binits{R.}},
\bauthor{\bsnm{Sharma}, \binits{A.S.}},
\bauthor{\bsnm{Tropp}, \binits{J.A.}},
\bauthor{\bsnm{McKeon}, \binits{B.J.}}:
\batitle{Model-based scaling of the streamwise energy density in high-reynolds-number turbulent channels}.
\bjtitle{Journal of Fluid Mechanics}
\bvolume{734},
\bfpage{275}--\blpage{316}
(\byear{2013})
\end{barticle}
\endbibitem

\bibitem[\protect\citeauthoryear{Abreu et~al.}{2020}]{abreu2020spectral}
\begin{barticle}
\bauthor{\bsnm{Abreu}, \binits{L.I.}},
\bauthor{\bsnm{Cavalieri}, \binits{A.V.G.}},
\bauthor{\bsnm{Schlatter}, \binits{P.}},
\bauthor{\bsnm{Vinuesa}, \binits{R.}},
\bauthor{\bsnm{Henningson}, \binits{D.S.}}:
\batitle{Spectral proper orthogonal decomposition and resolvent analysis of near-wall coherent structures in turbulent pipe flows}.
\bjtitle{Journal of Fluid Mechanics}
\bvolume{900},
\bfpage{11}
(\byear{2020})
\end{barticle}
\endbibitem

\bibitem[\protect\citeauthoryear{Gomez and McKeon}{2025}]{gomez2025linearAPGTBL}
\begin{barticle}
\bauthor{\bsnm{Gomez}, \binits{S.R.}},
\bauthor{\bsnm{McKeon}, \binits{B.J.}}:
\batitle{Linear analysis characterizes pressure gradient history effects in turbulent boundary layers}.
\bjtitle{Journal of Fluid Mechanics}
\bvolume{1002},
\bfpage{20}
(\byear{2025})
\end{barticle}
\endbibitem

\bibitem[\protect\citeauthoryear{Moarref et~al.}{2014}]{moarref2014low_order_represenanation}
\begin{botherref}
\oauthor{\bsnm{Moarref}, \binits{R.}},
\oauthor{\bsnm{Jovanovi{\'c}}, \binits{M.R.}},
\oauthor{\bsnm{Tropp}, \binits{J.A.}},
\oauthor{\bsnm{Sharma}, \binits{A.S.}},
\oauthor{\bsnm{McKeon}, \binits{B.J.}}:
A low-order decomposition of turbulent channel flow via resolvent analysis and convex optimization.
Physics of Fluids
\textbf{26}(5)
(2014)
\end{botherref}
\endbibitem

\bibitem[\protect\citeauthoryear{Towne et~al.}{2018}]{towne2018spectral}
\begin{barticle}
\bauthor{\bsnm{Towne}, \binits{A.}},
\bauthor{\bsnm{Schmidt}, \binits{O.T.}},
\bauthor{\bsnm{Colonius}, \binits{T.}}:
\batitle{Spectral proper orthogonal decomposition and its relationship to dynamic mode decomposition and resolvent analysis}.
\bjtitle{Journal of Fluid Mechanics}
\bvolume{847},
\bfpage{821}--\blpage{867}
(\byear{2018})
\end{barticle}
\endbibitem

\bibitem[\protect\citeauthoryear{Beneddine et~al.}{2017}]{beneddine2017unsteady}
\begin{barticle}
\bauthor{\bsnm{Beneddine}, \binits{S.}},
\bauthor{\bsnm{Yegavian}, \binits{R.}},
\bauthor{\bsnm{Sipp}, \binits{D.}},
\bauthor{\bsnm{Leclaire}, \binits{B.}}:
\batitle{Unsteady flow dynamics reconstruction from mean flow and point sensors: an experimental study}.
\bjtitle{Journal of Fluid Mechanics}
\bvolume{824},
\bfpage{174}--\blpage{201}
(\byear{2017})
\end{barticle}
\endbibitem

\bibitem[\protect\citeauthoryear{He et~al.}{2019}]{he2019data}
\begin{botherref}
\oauthor{\bsnm{He}, \binits{C.}},
\oauthor{\bsnm{Liu}, \binits{Y.}},
\oauthor{\bsnm{Gan}, \binits{L.}},
\oauthor{\bsnm{Lesshafft}, \binits{L.}}:
Data assimilation and resolvent analysis of turbulent flow behind a wall-proximity rib.
Physics of Fluids
\textbf{31}(2)
(2019)
\end{botherref}
\endbibitem

\bibitem[\protect\citeauthoryear{Lesshafft et~al.}{2019}]{lesshafft2019resolvent}
\begin{barticle}
\bauthor{\bsnm{Lesshafft}, \binits{L.}},
\bauthor{\bsnm{Semeraro}, \binits{O.}},
\bauthor{\bsnm{Jaunet}, \binits{V.}},
\bauthor{\bsnm{Cavalieri}, \binits{A.V.}},
\bauthor{\bsnm{Jordan}, \binits{P.}}:
\batitle{Resolvent-based modeling of coherent wave packets in a turbulent jet}.
\bjtitle{Physical Review Fluids}
\bvolume{4}(\bissue{6}),
\bfpage{063901}
(\byear{2019})
\end{barticle}
\endbibitem

\bibitem[\protect\citeauthoryear{Symon et~al.}{2019}]{symon2019tale}
\begin{barticle}
\bauthor{\bsnm{Symon}, \binits{S.}},
\bauthor{\bsnm{Sipp}, \binits{D.}},
\bauthor{\bsnm{McKeon}, \binits{B.J.}}:
\batitle{A tale of two airfoils: resolvent-based modelling of an oscillator versus an amplifier from an experimental mean}.
\bjtitle{Journal of Fluid Mechanics}
\bvolume{881},
\bfpage{51}--\blpage{83}
(\byear{2019})
\end{barticle}
\endbibitem

\bibitem[\protect\citeauthoryear{Preskett et~al.}{2024}]{preskett2024CTRSP}
\begin{bchapter}
\bauthor{\bsnm{Preskett}, \binits{T.}},
\bauthor{\bsnm{Jaiswal}, \binits{P.}},
\bauthor{\bsnm{Ganapathisubramani}, \binits{B.}},
\bauthor{\bsnm{Jaroslawski}, \binits{T.}},
\bauthor{\bsnm{Gomez}, \binits{S.R.}},
\bauthor{\bsnm{Vijay}, \binits{S.}},
\bauthor{\bsnm{McKeon}, \binits{B.J.}}:
\bctitle{Resolvent analysis of high reynolds number turbulent boundary layers subjected to pressure gradient histories}.
In: \bbtitle{Proceedings of the Summer Program},
p. \bfpage{43}
(\byear{2024})
\end{bchapter}
\endbibitem

\bibitem[\protect\citeauthoryear{Chavarin et~al.}{2020}]{chavarin2020resolvent}
\begin{barticle}
\bauthor{\bsnm{Chavarin}, \binits{A.}},
\bauthor{\bsnm{Efstathiou}, \binits{C.}},
\bauthor{\bsnm{Vijay}, \binits{S.}},
\bauthor{\bsnm{Luhar}, \binits{M.}}:
\batitle{Resolvent-based design and experimental testing of porous materials for passive turbulence control}.
\bjtitle{International Journal of Heat and Fluid Flow}
\bvolume{86},
\bfpage{108722}
(\byear{2020})
\end{barticle}
\endbibitem

\bibitem[\protect\citeauthoryear{Boutilier and Yarusevych}{2013}]{boutilier2013sensitivity}
\begin{barticle}
\bauthor{\bsnm{Boutilier}, \binits{M.S.}},
\bauthor{\bsnm{Yarusevych}, \binits{S.}}:
\batitle{Sensitivity of linear stability analysis of measured separated shear layers}.
\bjtitle{European Journal of Mechanics-B/Fluids}
\bvolume{37},
\bfpage{129}--\blpage{142}
(\byear{2013})
\end{barticle}
\endbibitem

\bibitem[\protect\citeauthoryear{Trefethen}{1999}]{trefethen1999spectra}
\begin{bbook}
\bauthor{\bsnm{Trefethen}, \binits{L.N.}}:
\bbtitle{Spectra and Pseudospectra: The Behaviour of Non-normal Matrices and Operators}.
\bpublisher{Springer}, \blocation{???}
(\byear{1999})
\end{bbook}
\endbibitem

\bibitem[\protect\citeauthoryear{de~Pando et~al.}{2014}]{de2014parametric}
\begin{bchapter}
\bauthor{\bsnm{Pando}, \binits{M.F.}},
\bauthor{\bsnm{Schmid}, \binits{P.J.}},
\bauthor{\bsnm{Lele}, \binits{S.}}:
\bctitle{Parametric sensitivity for large-scale aeroacoustic flows}.
In: \bbtitle{Proceedings of the Summer Program},
p. \bfpage{365}
(\byear{2014})
\end{bchapter}
\endbibitem

\bibitem[\protect\citeauthoryear{Fosas~de Pando and Schmid}{2017}]{fosas2017optimal}
\begin{barticle}
\bauthor{\bsnm{Pando}, \binits{M.}},
\bauthor{\bsnm{Schmid}, \binits{P.J.}}:
\batitle{Optimal frequency-response sensitivity of compressible flow over roughness elements}.
\bjtitle{Journal of Turbulence}
\bvolume{18}(\bissue{4}),
\bfpage{338}--\blpage{351}
(\byear{2017})
\end{barticle}
\endbibitem

\bibitem[\protect\citeauthoryear{Skene and Schmid}{2019}]{skene2019adjoint}
\begin{barticle}
\bauthor{\bsnm{Skene}, \binits{C.S.}},
\bauthor{\bsnm{Schmid}, \binits{P.J.}}:
\batitle{Adjoint-based parametric sensitivity analysis for swirling m-flames}.
\bjtitle{Journal of Fluid Mechanics}
\bvolume{859},
\bfpage{516}--\blpage{542}
(\byear{2019})
\end{barticle}
\endbibitem

\bibitem[\protect\citeauthoryear{Gomez et~al.}{2022}]{gomezCTRSP2022}
\begin{botherref}
\oauthor{\bsnm{Gomez}, \binits{S.R.}},
\oauthor{\bsnm{Williams}, \binits{C.T.}},
\oauthor{\bsnm{{Di Renzo}}, \binits{M.}},
\oauthor{\bsnm{Schmid}, \binits{P.J.}},
\oauthor{\bsnm{{McKeon}}, \binits{B.J.}}:
Adaptive resolvent analysis with application to high enthalpy flows.
Proceedings of the CTR Summer Program,
87--96
(2022)
\end{botherref}
\endbibitem

\bibitem[\protect\citeauthoryear{Hutchins et~al.}{2009}]{hutchins2009hot}
\begin{barticle}
\bauthor{\bsnm{Hutchins}, \binits{N.}},
\bauthor{\bsnm{Nickels}, \binits{T.B.}},
\bauthor{\bsnm{Marusic}, \binits{I.}},
\bauthor{\bsnm{Chong}, \binits{M.}}:
\batitle{Hot-wire spatial resolution issues in wall-bounded turbulence}.
\bjtitle{Journal of Fluid Mechanics}
\bvolume{635},
\bfpage{103}--\blpage{136}
(\byear{2009})
\end{barticle}
\endbibitem

\bibitem[\protect\citeauthoryear{Durst et~al.}{2001}]{durst2001situ}
\begin{barticle}
\bauthor{\bsnm{Durst}, \binits{F.}},
\bauthor{\bsnm{Zanoun}, \binits{E.-S.}},
\bauthor{\bsnm{Pashtrapanska}, \binits{M.}}:
\batitle{In situ calibration of hot wires close to highly heat-conducting walls}.
\bjtitle{Experiments in Fluids}
\bvolume{31}(\bissue{1}),
\bfpage{103}--\blpage{110}
(\byear{2001})
\end{barticle}
\endbibitem

\bibitem[\protect\citeauthoryear{{\"O}rl{\"u} et~al.}{2010}]{orlu2010near}
\begin{barticle}
\bauthor{\bsnm{{\"O}rl{\"u}}, \binits{R.}},
\bauthor{\bsnm{Fransson}, \binits{J.H.}},
\bauthor{\bsnm{Alfredsson}, \binits{P.H.}}:
\batitle{On near wall measurements of wall bounded flows—the necessity of an accurate determination of the wall position}.
\bjtitle{Progress in Aerospace Sciences}
\bvolume{46}(\bissue{8}),
\bfpage{353}--\blpage{387}
(\byear{2010})
\end{barticle}
\endbibitem

\bibitem[\protect\citeauthoryear{Bruun}{1996}]{bruun1996hot}
\begin{barticle}
\bauthor{\bsnm{Bruun}, \binits{H.H.}}:
\batitle{Hot-wire anemometry: principles and signal analysis}.
\bjtitle{Measurement Science and Technology}
\bvolume{7}(\bissue{10}),
\bfpage{024}
(\byear{1996})
\end{barticle}
\endbibitem

\bibitem[\protect\citeauthoryear{Ligrani and Bradshaw}{1987}]{ligrani1987spatial}
\begin{barticle}
\bauthor{\bsnm{Ligrani}, \binits{P.}},
\bauthor{\bsnm{Bradshaw}, \binits{P.}}:
\batitle{Spatial resolution and measurement of turbulence in the viscous sublayer using subminiature hot-wire probes}.
\bjtitle{Experiments in Fluids}
\bvolume{5}(\bissue{6}),
\bfpage{407}--\blpage{417}
(\byear{1987})
\end{barticle}
\endbibitem

\bibitem[\protect\citeauthoryear{Bhatia et~al.}{1982}]{bhatia1982corrections}
\begin{barticle}
\bauthor{\bsnm{Bhatia}, \binits{J.C.}},
\bauthor{\bsnm{Durst}, \binits{F.}},
\bauthor{\bsnm{Jovanovic}, \binits{J.}}:
\batitle{Corrections of hot-wire anemometer measurements near walls}.
\bjtitle{Journal of Fluid Mechanics}
\bvolume{122},
\bfpage{411}--\blpage{431}
(\byear{1982})
\end{barticle}
\endbibitem

\bibitem[\protect\citeauthoryear{Clauser}{1956}]{clauser1956turbulent}
\begin{barticle}
\bauthor{\bsnm{Clauser}, \binits{F.H.}}:
\batitle{The turbulent boundary layer}.
\bjtitle{Advances in applied mechanics}
\bvolume{4},
\bfpage{1}--\blpage{51}
(\byear{1956})
\end{barticle}
\endbibitem

\bibitem[\protect\citeauthoryear{Musker}{1979}]{musker1979explicit}
\begin{barticle}
\bauthor{\bsnm{Musker}, \binits{A.J.}}:
\batitle{Explicit expression for the smooth wall velocity distribution in a turbulent boundary layer}.
\bjtitle{AIAA journal}
\bvolume{17}(\bissue{6}),
\bfpage{655}--\blpage{657}
(\byear{1979})
\end{barticle}
\endbibitem

\bibitem[\protect\citeauthoryear{Dixit and Ramesh}{2009}]{dixit2009determination}
\begin{barticle}
\bauthor{\bsnm{Dixit}, \binits{S.A.}},
\bauthor{\bsnm{Ramesh}, \binits{O.}}:
\batitle{Determination of skin friction in strong pressure-gradient equilibrium and near-equilibrium turbulent boundary layers}.
\bjtitle{Experiments in fluids}
\bvolume{47}(\bissue{6}),
\bfpage{1045}--\blpage{1058}
(\byear{2009})
\end{barticle}
\endbibitem

\bibitem[\protect\citeauthoryear{Nagib et~al.}{2007}]{nagib2007approach}
\begin{barticle}
\bauthor{\bsnm{Nagib}, \binits{H.M.}},
\bauthor{\bsnm{Chauhan}, \binits{K.A.}},
\bauthor{\bsnm{Monkewitz}, \binits{P.A.}}:
\batitle{Approach to an asymptotic state for zero pressure gradient turbulent boundary layers}.
\bjtitle{Philosophical Transactions of the Royal Society A: Mathematical, Physical and Engineering Sciences}
\bvolume{365}(\bissue{1852}),
\bfpage{755}--\blpage{770}
(\byear{2007})
\end{barticle}
\endbibitem

\bibitem[\protect\citeauthoryear{Fernholz et~al.}{1996}]{fernholz1996new}
\begin{barticle}
\bauthor{\bsnm{Fernholz}, \binits{H.}},
\bauthor{\bsnm{Janke}, \binits{G.}},
\bauthor{\bsnm{Schober}, \binits{M.}},
\bauthor{\bsnm{Wagner}, \binits{P.}},
\bauthor{\bsnm{Warnack}, \binits{D.}}:
\batitle{New developments and applications of skin-friction measuring techniques}.
\bjtitle{Measurement Science and Technology}
\bvolume{7}(\bissue{10}),
\bfpage{1396}
(\byear{1996})
\end{barticle}
\endbibitem

\bibitem[\protect\citeauthoryear{Patel}{1965}]{patel1965calibration}
\begin{barticle}
\bauthor{\bsnm{Patel}, \binits{V.}}:
\batitle{Calibration of the preston tube and limitations on its use in pressure gradients}.
\bjtitle{Journal of Fluid Mechanics}
\bvolume{23}(\bissue{1}),
\bfpage{185}--\blpage{208}
(\byear{1965})
\end{barticle}
\endbibitem

\bibitem[\protect\citeauthoryear{K{\"a}hler et~al.}{2012}]{kahler2012uncertainty}
\begin{barticle}
\bauthor{\bsnm{K{\"a}hler}, \binits{C.J.}},
\bauthor{\bsnm{Scharnowski}, \binits{S.}},
\bauthor{\bsnm{Cierpka}, \binits{C.}}:
\batitle{On the uncertainty of digital piv and ptv near walls}.
\bjtitle{Experiments in fluids}
\bvolume{52}(\bissue{6}),
\bfpage{1641}--\blpage{1656}
(\byear{2012})
\end{barticle}
\endbibitem

\bibitem[\protect\citeauthoryear{Fuchs et~al.}{2023}]{fuchs2023wall}
\begin{barticle}
\bauthor{\bsnm{Fuchs}, \binits{T.}},
\bauthor{\bsnm{Bross}, \binits{M.}},
\bauthor{\bsnm{K{\"a}hler}, \binits{C.J.}}:
\batitle{Wall-shear-stress measurements using volumetric $\mu$ptv}.
\bjtitle{Experiments in Fluids}
\bvolume{64}(\bissue{6}),
\bfpage{115}
(\byear{2023})
\end{barticle}
\endbibitem

\bibitem[\protect\citeauthoryear{Huck et~al.}{2025}]{huck2025near}
\begin{barticle}
\bauthor{\bsnm{Huck}, \binits{P.D.}},
\bauthor{\bsnm{Yamakaitis}, \binits{M.J.}},
\bauthor{\bsnm{Fort}, \binits{C.}},
\bauthor{\bsnm{Bardet}, \binits{P.M.}}:
\batitle{Near-wall volumetric molecular tagging velocimetry with a fourier integral microscope}.
\bjtitle{Experiments in Fluids}
\bvolume{66}(\bissue{8}),
\bfpage{152}
(\byear{2025})
\end{barticle}
\endbibitem

\bibitem[\protect\citeauthoryear{Piomelli and Balaras}{2002}]{piomelli2002wall}
\begin{barticle}
\bauthor{\bsnm{Piomelli}, \binits{U.}},
\bauthor{\bsnm{Balaras}, \binits{E.}}:
\batitle{Wall-layer models for large-eddy simulations}.
\bjtitle{Annual review of fluid mechanics}
\bvolume{34}(\bissue{1}),
\bfpage{349}--\blpage{374}
(\byear{2002})
\end{barticle}
\endbibitem

\bibitem[\protect\citeauthoryear{Kawai and Larsson}{2012}]{kawai2012wall}
\begin{botherref}
\oauthor{\bsnm{Kawai}, \binits{S.}},
\oauthor{\bsnm{Larsson}, \binits{J.}}:
Wall-modeling in large eddy simulation: Length scales, grid resolution, and accuracy.
Physics of fluids
\textbf{24}(1)
(2012)
\end{botherref}
\endbibitem

\bibitem[\protect\citeauthoryear{Park and Moin}{2014}]{park2014improved}
\begin{botherref}
\oauthor{\bsnm{Park}, \binits{G.I.}},
\oauthor{\bsnm{Moin}, \binits{P.}}:
An improved dynamic non-equilibrium wall-model for large eddy simulation.
Physics of Fluids
\textbf{26}(1)
(2014)
\end{botherref}
\endbibitem

\bibitem[\protect\citeauthoryear{Gomez}{2024}]{gomez2024linear}
\begin{botherref}
\oauthor{\bsnm{Gomez}, \binits{S.R.}}:
Linear amplification in nonequilibrium turbulent boundary layers.
PhD thesis,
California Institute of Technology
(2024)
\end{botherref}
\endbibitem

\bibitem[\protect\citeauthoryear{Mattsson and Nordstr{\"o}m}{2004}]{mattsson2004SBP}
\begin{barticle}
\bauthor{\bsnm{Mattsson}, \binits{K.}},
\bauthor{\bsnm{Nordstr{\"o}m}, \binits{J.}}:
\batitle{Summation by parts operators for finite difference approximations of second derivatives}.
\bjtitle{Journal of Computational Physics}
\bvolume{199}(\bissue{2}),
\bfpage{503}--\blpage{540}
(\byear{2004})
\end{barticle}
\endbibitem

\bibitem[\protect\citeauthoryear{Malik}{1990}]{malik1990numerical}
\begin{barticle}
\bauthor{\bsnm{Malik}, \binits{M.R.}}:
\batitle{Numerical methods for hypersonic boundary layer stability}.
\bjtitle{Journal of Computational Physics}
\bvolume{86}(\bissue{2}),
\bfpage{376}--\blpage{413}
(\byear{1990})
\end{barticle}
\endbibitem

\bibitem[\protect\citeauthoryear{Schlatter and {\"O}rl{\"u}}{2010}]{schlatter2010assessment}
\begin{barticle}
\bauthor{\bsnm{Schlatter}, \binits{P.}},
\bauthor{\bsnm{{\"O}rl{\"u}}, \binits{R.}}:
\batitle{Assessment of direct numerical simulation data of turbulent boundary layers}.
\bjtitle{Journal of Fluid Mechanics}
\bvolume{659},
\bfpage{116}--\blpage{126}
(\byear{2010})
\end{barticle}
\endbibitem

\bibitem[\protect\citeauthoryear{Pope}{2000}]{pope2000turbulent}
\begin{bbook}
\bauthor{\bsnm{Pope}, \binits{S.B.}}:
\bbtitle{Turbulent Flows}.
\bpublisher{Cambridge University Press}, \blocation{???}
(\byear{2000})
\end{bbook}
\endbibitem

\bibitem[\protect\citeauthoryear{Hoyas and Jim{\'e}nez}{2006}]{hoyas2006scaling}
\begin{barticle}
\bauthor{\bsnm{Hoyas}, \binits{S.}},
\bauthor{\bsnm{Jim{\'e}nez}, \binits{J.}}:
\batitle{Scaling of the velocity fluctuations in turbulent channels up to re $\tau$= 2003}.
\bjtitle{Physics of Fluids}
\bvolume{18}(\bissue{1}),
\bfpage{011702}
(\byear{2006})
\end{barticle}
\endbibitem

\bibitem[\protect\citeauthoryear{Pozuelo et~al.}{2022}]{pozuelo2022adverse}
\begin{barticle}
\bauthor{\bsnm{Pozuelo}, \binits{R.}},
\bauthor{\bsnm{Li}, \binits{Q.}},
\bauthor{\bsnm{Schlatter}, \binits{P.}},
\bauthor{\bsnm{Vinuesa}, \binits{R.}}:
\batitle{An adverse-pressure-gradient turbulent boundary layer with nearly constant $\beta \simeq 1.4$ up to $re_{\theta}\simeq 8,700$}.
\bjtitle{Journal of Fluid Mechanics}
\bvolume{939},
\bfpage{34}
(\byear{2022})
\end{barticle}
\endbibitem

\bibitem[\protect\citeauthoryear{Eitel-Amor et~al.}{2014}]{eitel2014simulationZPG}
\begin{barticle}
\bauthor{\bsnm{Eitel-Amor}, \binits{G.}},
\bauthor{\bsnm{{\"O}rl{\"u}}, \binits{R.}},
\bauthor{\bsnm{Schlatter}, \binits{P.}}:
\batitle{Simulation and validation of a spatially evolving turbulent boundary layer up to $re_\theta= 8300$}.
\bjtitle{International Journal of Heat and Fluid Flow}
\bvolume{47},
\bfpage{57}--\blpage{69}
(\byear{2014})
\end{barticle}
\endbibitem

\bibitem[\protect\citeauthoryear{Benedict and Gould}{1996}]{benedict1996towards}
\begin{barticle}
\bauthor{\bsnm{Benedict}, \binits{L.}},
\bauthor{\bsnm{Gould}, \binits{R.}}:
\batitle{Towards better uncertainty estimates for turbulence statistics}.
\bjtitle{Experiments in fluids}
\bvolume{22}(\bissue{2}),
\bfpage{129}--\blpage{136}
(\byear{1996})
\end{barticle}
\endbibitem

\bibitem[\protect\citeauthoryear{Chu}{1965}]{chu1965energy}
\begin{barticle}
\bauthor{\bsnm{Chu}, \binits{B.-T.}}:
\batitle{On the energy transfer to small disturbances in fluid flow (part i)}.
\bjtitle{Acta Mechanica}
\bvolume{1}(\bissue{3}),
\bfpage{215}--\blpage{234}
(\byear{1965})
\end{barticle}
\endbibitem

\end{thebibliography}

\end{document}